\documentclass[preprint, review, 3p, 12pt]{elsarticle}
\usepackage[american]{babel}
\usepackage[T1]{fontenc}
\usepackage{amsmath}
\usepackage{amssymb}
\usepackage{mathtools}
\usepackage{siunitx}
\usepackage{miller}
\usepackage{booktabs}
\usepackage{dcolumn}
\usepackage{makecell}
\usepackage{xcolor}
\usepackage{enumitem}
\usepackage{subfig}
\usepackage{csquotes}
\usepackage{soul}
\usepackage{comment}
\usepackage{pdfpages}
\usepackage{setspace}
\usepackage{url}
\usepackage{varioref}
\usepackage{hyperref}
\usepackage[nameinlink]{cleveref}
\newcolumntype{d}[1]{D..{#1}}

\definecolor{tud0d}{RGB}{83,83,83}
\definecolor{tud0c}{RGB}{137,137,137}
\definecolor{tud0b}{RGB}{181,181,181}
\definecolor{tud0a}{RGB}{220,220,220}

\definecolor{tud1a}{RGB}{93,133,195}
\definecolor{tud2a}{RGB}{0,156,218}
\definecolor{tud3a}{RGB}{80,182,149}
\definecolor{tud4a}{RGB}{175,204,80}
\definecolor{tud5a}{RGB}{221,223,72}
\definecolor{tud6a}{RGB}{255,224,92}
\definecolor{tud7a}{RGB}{248,186,60}
\definecolor{tud8a}{RGB}{238,122,52}
\definecolor{tud9a}{RGB}{233,80,62}
\definecolor{tud10a}{RGB}{201,48,142}
\definecolor{tud11a}{RGB}{128,69,151}

\definecolor{tud1b}{RGB}{0,90,169}
\definecolor{tud2b}{RGB}{0,131,204}
\definecolor{tud3b}{RGB}{0,157,129}
\definecolor{tud4b}{RGB}{153,192,0}
\definecolor{tud5b}{RGB}{201,212,0}
\definecolor{tud6b}{RGB}{253,202,0}
\definecolor{tud7b}{RGB}{245,163,0}
\definecolor{tud8b}{RGB}{236,101,0}
\definecolor{tud9b}{RGB}{230,0,26}
\definecolor{tud10b}{RGB}{166,0,132}
\definecolor{tud11b}{RGB}{114,16,133}

\definecolor{tud1c}{RGB}{0,78,138}
\definecolor{tud2c}{RGB}{0,104,157}
\definecolor{tud3c}{RGB}{0,136,119}
\definecolor{tud4c}{RGB}{127,171,22}
\definecolor{tud5c}{RGB}{177,189,0}
\definecolor{tud6c}{RGB}{215,172,0}
\definecolor{tud7c}{RGB}{210,135,0}
\definecolor{tud8c}{RGB}{204,76,3}
\definecolor{tud9c}{RGB}{185,15,34}
\definecolor{tud10c}{RGB}{149,17,105}
\definecolor{tud11c}{RGB}{97,28,115}

\definecolor{tud1d}{RGB}{36,53,114}
\definecolor{tud2d}{RGB}{0,78,115}
\definecolor{tud3d}{RGB}{0,113,94}
\definecolor{tud4d}{RGB}{106,139,55}
\definecolor{tud5d}{RGB}{153,166,4}
\definecolor{tud6d}{RGB}{174,142,0}
\definecolor{tud7d}{RGB}{190,111,0}
\definecolor{tud8d}{RGB}{169,73,19}
\definecolor{tud9d}{RGB}{156,28,38}
\definecolor{tud10d}{RGB}{115,32,84}
\definecolor{tud11d}{RGB}{76,34,106}
\usepackage{transparent}
\usepackage{tikz}
\usepackage{pgfplots}
\usepackage{pgfplotstable}
\pgfplotsset{compat=newest}
\pgfplotsset{
        layers/my layer set/.define layer set={
            background,
            main,
            foreground
        }{},
        set layers=my layer set,
    }
\usepgfplotslibrary{external}
\tikzsetexternalprefix{img/tikz/}
\usepgfplotslibrary{groupplots}
\usepgfplotslibrary{fillbetween}
\usepgfplotslibrary{colormaps}
\usetikzlibrary{automata}
\usetikzlibrary{positioning}
\usetikzlibrary{calc}
\usetikzlibrary{plotmarks}
\usetikzlibrary{arrows}
\tikzset{
    >=stealth',
    pil/.style={
           ->,
           thick,
           shorten <=2pt,
           shorten >=2pt,}
}

\usepackage[percent]{overpic}
\tikzset{external/system call={
    lualatex
    \tikzexternalcheckshellescape -halt-on-error -interaction=batchmode
        -jobname "\image" "\texsource"
        }}
\pgfkeys{/pgf/number format/.cd,
    1000 sep={\,},
    min exponent for 1000 sep=4}
\pgfplotsset{unit code/.code 2 args={\si{#1#2}}}
\pgfplotsset{
    legend image with text/.style={
        legend image code/.code={%
            \node[anchor=center] at (0.3cm,0cm) {#1};
        }
    },
}
\biboptions{sort&compress}
\DeclareSIUnit\angstrom{\text{\AA}}
\hypersetup{
pdfborder={0 0 0},
breaklinks=true,}

\def\doubleunderline#1{\underline{\underline{#1}}}

\newcommand{\qty}{\SI}

\begin{document}

\begin{frontmatter}

    \title{The nature of deformation-induced dislocations in SrTiO\textsubscript{3}: \newline Insights from atomistic simulations}
    
    \author[1]{Arne J. Klomp\corref{cor1}}
    \ead{klomp@mm.tu-darmstadt.de}
    \author[2]{Lukas Porz}
    \author[1]{Karsten Albe}

    \cortext[cor1]{Corresponding author}
    \address[1]{Materials Modelling Division, Technical University of Darmstadt, Otto-Berndt-Str. 3, 64287 Darmstadt, Germany}
    \address[2]{Nonmetallic-Inorganic Materials Group, Technical University of Darmstadt, Alarich-Weiss-Str. 3, 64287 Darmstadt, Germany}

    \begin{abstract}
        
        The nature of mechanically induced dislocations in SrTiO\textsubscript{3} at low temperatures has been a disputed matter for a long time.
        Here we  provide a systematic overview of the existing knowledge on dislocations in stoichiometric SrTiO\textsubscript{3} complemented by computational models of several of the proposed dislocation types.
        Based on atomistic simulations we reveal their structure and mobility and put our focus on the splitting  into partial dislocations, because this mechanism is held responsible for the good dislocation mobility in strontium titanate.
        Our results reveal that dislocations with a \hkl{1-10} glide plane (types $\mathcal{A}$, $\mathcal{B}$, and $\mathcal{C}$) show easy glide behavior due to their glide dissociated configuration.
        The motion of dislocations on \hkl{001} glide planes (type $\mathcal{E}$), however, is prohibited  due to high Peierls barriers, as is the case for the \hkl{1-1-2} glide plane (type $\mathcal{D}$).
        For dislocations with (partial) edge character the dissociation into partials is shown to be very sensitive to the charge state of the dislocation core.
        Since, in turn, dislocation mobility is strongly related to the dissociation of dislocations, the macroscopic ductility may in some cases have a direct relation to the charge at the dislocation core.


    \end{abstract}
    
    \begin{keyword}
        dislocations \sep plasticity \sep strontium titanate \sep perovskite \sep molecular dynamics
    \end{keyword}
    
\end{frontmatter}


\section{Introduction}

    After the discovery of low temperature ductility in strontium titanate single crystals (SrTiO\textsubscript{3} or STO) about 20 years ago, the interest in dislocation behavior in ceramics has re-surged \cite{Brunner2001}, since it opened new avenues for  dislocation engineering of functional ceramics \cite{Merkle2003, Merkle2008}.
    Once dislocations can be controlled reliably \cite{Porz2021}, new and improved functionality and increased resistance against brittle fracture can be expected for functional ceramics.
    This, however, requires a detailed understanding of dislocation configurations and mobility on the atomic level.
    The cubic perovskite structure of SrTiO\textsubscript{3} serves as a prototype system for the class of perovskite ceramics which bears significant technological importance, e.g. for applications related to (anti-)ferroelectricity and superconductivity \cite{Bhalla2000}.
    Dislocations in SrTiO\textsubscript{3} can be easily introduced into single crystals by mechanical force \cite{Porz2021}.
    However, despite the wealth of literature already published on this topic, the nature of mechanically introduced dislocations at low temperatures is not agreed upon.
 
    Dislocations in ionic crystals have been investigated for almost as long as dislocations in metals \cite{HardouinDuparc2017}.
    Yet, the crystallographic variety and complexity of ionic compounds has led to much fewer studies in this field compared to dislocations in metals.
    As comprehensively discussed by Eshelby \emph{et al.} \cite{Eshelby1958} and Whitworth \cite{Whitworth1975} dislocations in ionic structures can be inherently charged, if the crystal structure allows Burgers vector cuts of edge dislocations in a non-charge balanced manner.
    These dislocations, therefore, affect the spatial distribution of charged point defects in their surroundings not only due to their elastic but also due to their electrostatic fields.
    Depending on the resulting charge distribution, a dislocation can be subjected to pinning or the charge on a dislocation can try to drive the dislocation away.
    Thus, the charge on a dislocation can have a measurable impact on the yield strength and be an important factor in the light of crystal plasticity.
    
    The electrostatic energy contribution from the unbalanced charge can be significant.
    To address this effect different schemes for maintaining charge neutrality in a macroscopic crystal while changing the charge on the dislocation line were devised {e.g. Ref.~\cite{Marrocchelli2015}}.
    In this context also kinetic aspects matter.
    Swiftly moving dislocations, for example cannot attain local  charge neutrality in an intrinsic crystal, if they escape from a cloud of slowly moving compensating defects.
    Static charged dislocations, in contrast, can eventually equilibrate over time, e.g., by changing the occupancy of atomic sites at the dislocation core.

    The role of the dissociation of large Burgers vector dislocations into partial dislocations, that is typical for large unit cells, is another important aspect, which has been discussed in depth by Mitchell \emph{et al.} \cite{Mitchell1985}.
    They noted that the condition of close packed planes and short Burgers vectors is broken to some degree in ceramic materials due to electrostatic repulsion of ions.
    For the same reason it was suspected that stable stacking faults in ceramics require at least one ionic sub-lattice remaining undisturbed.
    Focusing on the high temperature regime they stated: \textquote{{\em Ductile} ceramics have remained an elusive and perhaps impossible goal, although {\em toughened} ceramics have emerged as a reality}.
    This was largely attributed to dislocations climb dissociating as soon as temperature allows for diffusional transport.
    Climb dissociation is then favored over a glide dissociated dislocation for energetic reasons as outlined below.
    The transition from a glide dissociated dislocation to a climb dissociated dislocation would, however, need constriction of the glide dissociated dislocation and a considerable amount of diffusion.
    
    A further complication with dislocations arises in systems containing dopants or impurities.
    Extrinsic point defects can massively alter the defect equilibria, especially at low temperatures.
    Also, the configuration of dislocation cores can change during movement due to sweep-up processes.
    
    There is considerable uncertainty in the research community regarding the nature of mechanically induced dislocations in SrTiO\textsubscript{3} at low temperatures.
    Therefore, 
    in this study, we  provide a systematic overview of the existing knowledge on dislocations in stoichiometric SrTiO\textsubscript{3} complemented by computational models of several of the proposed dislocation types.
    Based on atomistic simulations we discuss their structure and properties and put a focus on the splitting into partial dislocations because this mechanism is held responsible for the good dislocation mobility in strontium titanate.

    \begin{figure*}[htb]
        \centering
        \onehalfspacing
            \subfloat[\label{fig:types_overview_A}][Dislocation type $\mathcal{A}$.]{
                \includegraphics[]{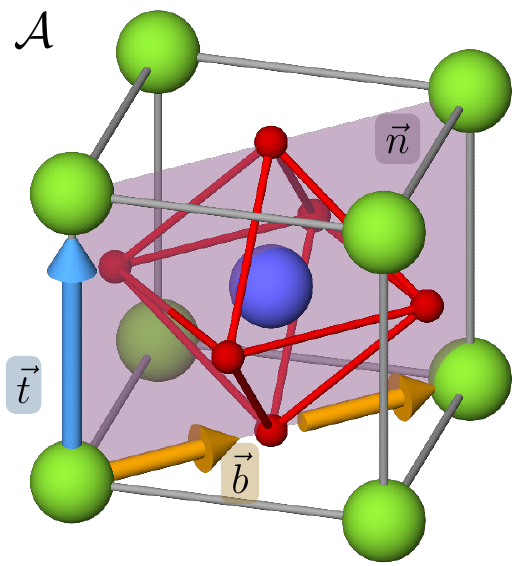}
            }\hfil
            \subfloat[\label{fig:types_overview_B}][Dislocation type $\mathcal{B}$.]{
                \includegraphics[]{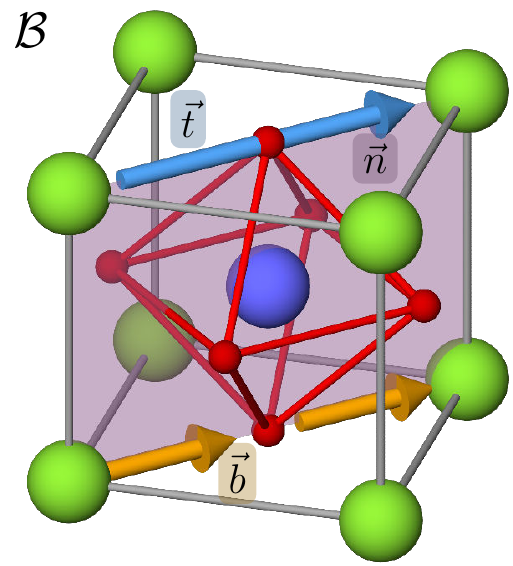}               
            }\hfil
            \subfloat[\label{fig:types_overview_C}][Dislocation type $\mathcal{C}$.]{
                \includegraphics[]{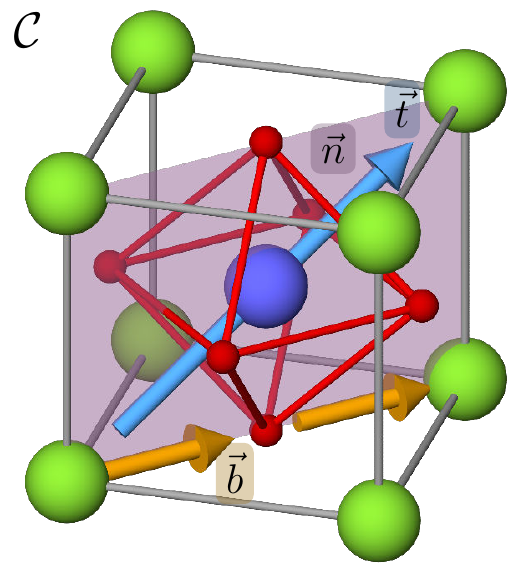}               
            }\hfil
            \subfloat[\label{fig:types_overview_D}][Dislocation type $\mathcal{D}$.]{
                \includegraphics[]{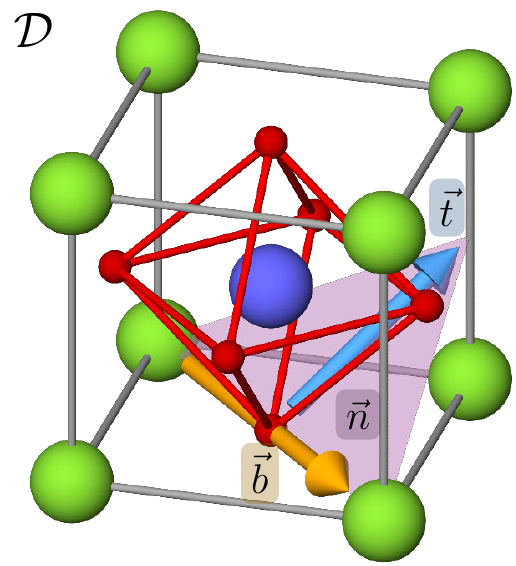}                
            }\hfil
            \subfloat[\label{fig:types_overview_E}][Dislocation type $\mathcal{E}$.]{
                \includegraphics[]{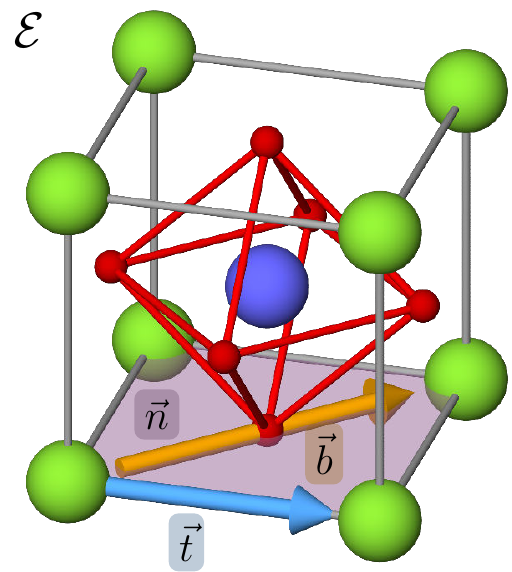}                
            }\hfil
            \subfloat[\label{tab:types_overview}][Overview of dislocation types and characteristic vectors.]{
                \begin{tabular}{r | c c c | l l l}
                    \toprule
                        & $\vec{b}$    & $\vec{t}$  & $\vec{n}$   & type  & possible configuration & Refs. \\
                    \midrule
                    $\mathcal{A}$ & $a\hkl<110>$ & \hkl<001>  & \hkl{1-10}  & edge  & full/glide/climb       & \cite{Taeri2004, Sigle2006,Hirel2012,Yang2011,Hirel2016,Castillo-Rodriguez2011,Yang2011a} \\
                    $\mathcal{B}$ & $a\hkl<110>$ & \hkl<110>  & \hkl{1-10}  & screw & full/glide             & \cite{Gumbsch2001,Taeri2004,Sigle2006,Castillo-Rodriguez2010,Hirel2012,Kondo2012,Castillo-Rodriguez2011} \\
                    $\mathcal{C}$ & $a\hkl<110>$ & \hkl<111>  & \hkl{1-10}  & mixed & full/glide             & \cite{Yang2011,Jin2013,Javaid2017} \\
                    $\mathcal{D}$ & $a\hkl<110>$ & \hkl<1-11> & \hkl{1-1-2} & edge  & full/climb             & \cite{Jin2013} \\
                    $\mathcal{E}$ & $a\hkl<110>$ & \hkl<100>  & \hkl{001}   & mixed & full/glide/climb       & \cite{Mao1996,Yang2009} \\ 
                    \bottomrule
                \end{tabular}
            }\hfil
        \caption{
            (a) to (e): Crystallographic orientation of defining vectors for all studied dislocation types $\mathcal{A}$ to $\mathcal{E}$.
            Unit cell of cubic SrTiO\textsubscript{3} with Burgers vector $\vec{b}$ (orange), line vector $\vec{t}$ (blue) and slip plane $\vec{n}$ (pink).
            The full Burgers vector has been split into two equal parts to indicate glide dissociation where applicable.
            Details are discussed in the respective sections.
            (f): Summary of the vectors corresponding to (a) to (e) and dislocation configurations mentioned in the given references.
        }
        \label{fig:types_overview}
    \end{figure*}

    An overview of the studied dislocation types is given in \Cref{fig:types_overview}.
    In this study, we specifically address the following points:
    \begin{itemize}
        \item The dislocation types that have been mostly studied so far are type $\mathcal{A}$ and $\mathcal{B}$ dislocations. We investigate the equilibrium configuration and splitting of these pure  edge and screw dislocations of the \hkl<110>\hkl{1-10} type.
        \item Type $\mathcal{C}$ dislocations have only been reported in some TEM studies \cite{Jin2013}. We clarify if these dislocations tend to glide- or climb-dissociate and if they remain as mixed dislocations or decompose into edge and screw component.
        \item It was proposed that a type $\mathcal{D}$ dislocation can form by reaction of two appropriate $\mathcal{C}$ dislocations \cite{Jin2013}. We investigate the equilibrium configuration of this dislocation and compare to type $\mathcal{A}$ dislocations where the line direction differs.
        \item Type $\mathcal{E}$ dislocations behave analog to the other dislocation types with $\vec{b}=\hkl<110>$ Burgers vector. We investigate the influence of the different glide plane.
    \end{itemize}

    In the next section we describe our computational methodology.
    Next, we give a short review on the literature regarding dislocation mechanics in SrTiO\textsubscript{3}.
    Then we turn to studying all the dislocation types consecutively and give answers to the above questions.

\section{Methods}

    \subsection{Density Functional Theory Calculations}
        
        For obtaining $\gamma$-surfaces \cite{Vitek1968} shown in \Cref{fig:sf_surface} we have applied density-functional theory calculations using \texttt{ABINIT} \cite{Gonze2016}.
        The electron density is modeled based on the projector augmented-wave method (PAW) with datasets created for the Perdew-Burke-Ernzerhof exchange-correlation functional using \texttt{ATOMPAW} (https://github.com/atompaw/atompaw); as they are provided by \texttt{ABINIT}.
        The unit cell volume has been relaxed to the equilibrium lattice parameters and convergence with respect to energy cutoff and reciprocal space mesh has been checked.
        Both calculated $\gamma$-surfaces are obtained with columnar 3D periodic simulation cells that are of size one unit cell in plane and of length six unit cells perpendicular to the stacking fault plane.
        This distance between two neighboring stacking faults is, thus, only three unit cells, which may hardly be called converged with respect to simulation cell size.
        Since we are interested in the general shape of the $\gamma$-surface and not in the exact stacking fault energies this inaccuracy is accepted.
        The sampling of the $\gamma$-surface is performed by shifting one half of the columnar cell on a regular 2D grid relative to the other part.
        While the cubic lattice parameter is held fixed in the in-plane directions, atomic positions as well as simulation cell size perpendicular to the stacking fault are carefully relaxed using the Broyden-Fletcher-Goldfarb-Shanno procedure.

    \subsection{Molecular Dynamics Simulations}
       
\begin{figure}[htb]
	    \centering
            \includegraphics[]{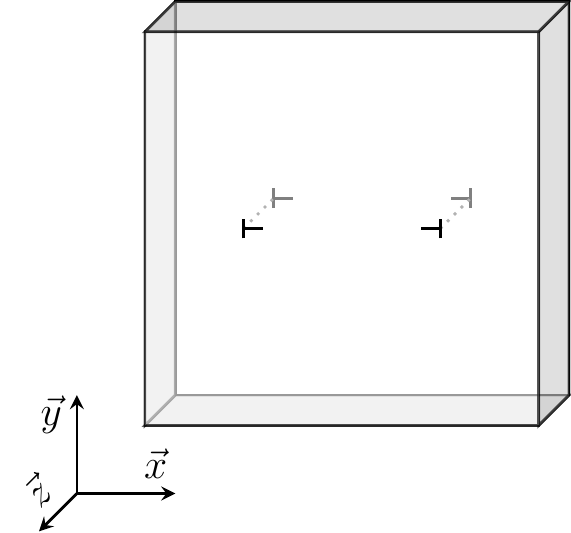}
            \caption{
                Schematic drawing of the simulation cell setup indicating the cell orientation and the dislocation lines.
                The crystallographic orientation corresponding to the directions $\vec{x}$, $\vec{y}$ and $\vec{z}$ for the different types of dislocations are given in \Cref{tab:simcell_directions}.
                Dislocation lines are always aligned with the $\vec{z}$-axis and all boundaries are fully periodic.
                }
            \label{fig:simcell_schematic}
        \end{figure}

        For investigating the structure and mobility of dislocations in SrTiO\textsubscript{3}, we have employed molecular dynamics simulations using \texttt{LAMMPS} \cite{Plimpton1995,Thompson2022}.   
        An interatomic potential by Thomas \emph{et al.} was used which provides a reliable description of lattice and defect properties \cite{Thomas2005} but contains fixed effective ionic charges.
        In order to ensure charge neutrality only  stoichiometric SrTiO\textsubscript{3} configurations have been studied.
        As model configurations, dislocation dipoles in orthogonal simulation cells were chosen, see \Cref{fig:simcell_schematic}.
        Each simulation cell has two long edges along $\vec{x}$ and $\vec{y}$ and one short edge along $\vec{z}$.
       
        The dislocation line vector \(\vec{t}\) is always aligned with the \(\vec{z}\)-direction of the simulation cell.
        If the dislocation type additionally has an edge component or is of edge type, the edge component is aligned with $\vec{y}$.
        The crystallographic directions that are aligned with the edges of the simulation cell are given in \Cref{tab:simcell_directions}.
        This way simulation cells containing between \num{390000} and \num{800000} atoms of cubic SrTiO\textsubscript{3} are created.

        \begin{table}[htb]
            \caption{Crystallographic axes that are aligned with the simulation cell directions $\vec{x}$, $\vec{y}$ and $\vec{z}$ as defined in the schematic drawing in \Cref{fig:simcell_schematic}. The $\vec{z}$ of the simulation cell (last column of this table) is always aligned with the dislocation line vector $\vec{t}$ (or a permutation thereof), see \Cref{fig:types_overview}~f.}
            \label{tab:simcell_directions}
            \begin{center}
            \begin{tabular}{r | c c c}
                    \toprule
                    & $\vec{x}$ & $\vec{y}$ & $\vec{z}$ \\
                    \midrule
                    $\mathcal{A}$ & \hkl[1-10] & \hkl[110] & \hkl[001]\\
                    $\mathcal{B}$ & \hkl[001] & \hkl[1-10] & \hkl[110]\\
                    $\mathcal{C}$ & \hkl[1-10] & \hkl[11-2] & \hkl[111]\\
                    $\mathcal{D}$ & \hkl[11-2] & \hkl[-110] & \hkl[111]\\
                    $\mathcal{E}$ & \hkl[100] & \hkl[010] & \hkl[001]\\
                    \bottomrule
            \end{tabular}
            \end{center}
        \end{table}

        Dislocations with edge contributions were constructed by deleting individual lines of atoms spanning half the simulation cell  (i.e. a ribbon in between the two dislocation lines).
        In doing so, one obtains two dislocations with oppositely oriented Burgers vectors.
        For the screw component the analytical continuum solution for the screw dislocation displacement field is applied to the atoms \cite{Cai2016}.

        The as-created dislocation configurations were  carefully relaxed.
        By step-wise removal of constraints to the atomic movement we avoided unphysical reconstruction of the dislocation cores.
        The energy of the configurations was  measured during NPT molecular dynamics runs at temperatures between \qty{20}{\kelvin} to \qty{50}{\kelvin} and zero applied pressure.
        Each dislocation configuration was simulated multiple times with different temperature initializations of the MD simulation to check for consistency.
        By investigating finite-size scaling of the results, we made sure that the reported results are independent of the simulated cell sizes.
        The software packages \texttt{ASE} and \texttt{Ovito} were employed for setting up the cells, applying atom displacements and post-processing \cite{HjorthLarsen2017, Stukowski2010}.

        Screw dislocations cannot possess a net charge, since atoms (or ions for that matter) are simply displaced along the dislocation line.
        However, as-prepared edge dislocation are not necessarily charge neutral because the introduced half-planes can be terminated non-stoichiometrically, i.e., the dislocations are inherently charged.
        For restoring the charge balance at the dislocation cores we shifted  oxygen ions from one dislocation core to the other since they are the most mobile species \cite{Marrocchelli2015}.
        Since charge compensation/balancing of the dislocation cores has a significant impact on the energy of the dislocations as well as their configurations, we also studied the influence of the amount of excess charge on the configuration of the dislocation.
        The results are described in the sections on dislocation types $\mathcal{A}$ and $\mathcal{C}$.

        Finally, for dislocation types $\mathcal{A}$, $\mathcal{B}$, and $\mathcal{C}$ (which we expect to be mobile) also mechanical loading was simulated.
        Loading is performed by ramping up the desired component of the barostat until the dislocation moves.
        To ensure that the calculated stress is not increased due to the dynamic loading, the samples are held at constant load just below this value for \qty{24}{\pico\second}.
        The loading simulations are performed in the athermal regime (at \qty{10}{\kelvin}) and also with large thermal activation (at \qty{1500}{\kelvin}).
        Additionally, we also conducted simulations where the z-direction was increased from 2~unit~cells to 60~unit~cells which allows dislocation motion by a kink formation mechanism.
        Based on Ref. \cite{Yang2011} we expect this size to be sufficient to see kink nucleation and propagation.

\section{Results}

    \subsection{Dislocation types in strontium titanate}
    
        In order to identify the relevant dislocations in SrTiO\textsubscript{3} we first review the existing experimental and theoretical literature.
        The dislocation types investigated in this publication are shown in \Cref{fig:types_overview}. 
        All of them possess a Burgers vector \hkl<110> but differ in the orientation of the dislocation line, i.e., they have different edge or screw character.
        The types $\mathcal{A}$, $\mathcal{B}$, and $\mathcal{C}$ are characterized by a \hkl{1-10} glide plane normal.
        Dislocation types $\mathcal{D}$ and $\mathcal{E}$ are also investigated here because they are either related to the former dislocation types by dislocation reactions (type $\mathcal{D}$) or because their Burgers vector suggests a relation to the dislocations of types $\mathcal{A}$ to $\mathcal{C}$ (type $\mathcal{E}$).

    \subsection{Dislocation splitting}
    
        Gumbsch {\em et al.} found that SrTiO\textsubscript{3} is ductile from \qty{110}{\kelvin} to \qty{1000}{\kelvin} by compression of single crystals along various crystallographic axes \cite{Brunner2001, Gumbsch2001, Taeri2004}.
        The ductility is attributed to dislocations with Burgers vector \hkl<110> on a slip plane \hkl{1-10}, i.e., type $\mathcal{A}$, $\mathcal{B}$, or $\mathcal{C}$ dislocations.
        In a first step, we will clarify why dislocations with \hkl<110> Burgers vector  play such an important role in SrTiO\textsubscript{3}. Dislocations reside on glide planes which have close packing and short Burgers vectors  $\vec{b}$. These Burgers vectors are \hkl<100>, \hkl<110>, and \hkl<111> in a cubic system.
        The elastic line energy $\Gamma_{\text{elast}}$ of a dislocation is
        proportional to the square of the Burgers vector $\vec{b}$ \cite{Cai2016}:
        
        \begin{align}
            \Gamma_{\text{elast}} \propto \left|\vec{b}\right|^2 = b^2 \text{.}\label{eq:elastic_energy_proportionality}
        \end{align}
        Comparing $b^2$ for the three considered options, we note that the elastic line energy of the dislocations doubles or triples for the non-primitive Burgers vectors \hkl<110> and \hkl<111>, respectively, see \Cref{tab:squared_b}.
        
        \begin{table}[ht] 
            \caption{Relevant slip systems in SrTiO\textsubscript{3} with the Burgers vector in angular backets and the slip plane normal in curly braces. The square of the Burgers vector $\left|\vec{b}\right|^2$ is computed using the unit cell parameter $a$. It is proportional to the elastic line energy of a dislocation, see Equation \eqref{eq:elastic_energy_proportionality}. A \hkl<110>\hkl{1-10} dislocation which is split into two equal collinear partials has the same elastic line energy as a full \hkl<100>\hkl{001} dislocation.}
            \label{tab:squared_b}
            \begin{center}
                \begin{tabular}{r l}
                    \toprule
                    dislocation & $\left|\vec{b}\right|^2$ \\
                    \midrule
                    \hkl<100>\hkl{001}  & $a^2$\\
                    \hkl<110>\hkl{001}  & $2 a^2$\\
                    \hkl<001>\hkl{1-10} & $a^2$\\
                    \hkl<110>\hkl{1-10} & $2 a^2$\\
                    2 x $\frac{1}{2} \hkl<110>\hkl{1-10}$ & 2 x $\frac{1}{2}a^2$\\
                    \bottomrule
                \end{tabular}
            \end{center}
        \end{table}
        
        Since dislocations with \hkl<111> Burgers vector are unfeasible because of their high elastic energies \cite{Klomp2018}, we can restrict ourselves to glide systems with \hkl<100> and \hkl<110> Burgers vectors.
        Regarding the glide planes, dislocations tend to favor densely packed planes.
        The \hkl{111} planes have the loosest packing and, therefore, don't play a role.
        In contrast glide planes  \hkl{100} and \hkl{110} have a denser packing and are likely to enable easy glide.
        Thus, we are left with the two Burgers vectors \hkl<100> and \hkl<110> and the two glide planes \hkl{100} and \hkl{110}.
        
\begin{figure*}
            \centering
            \includegraphics[]{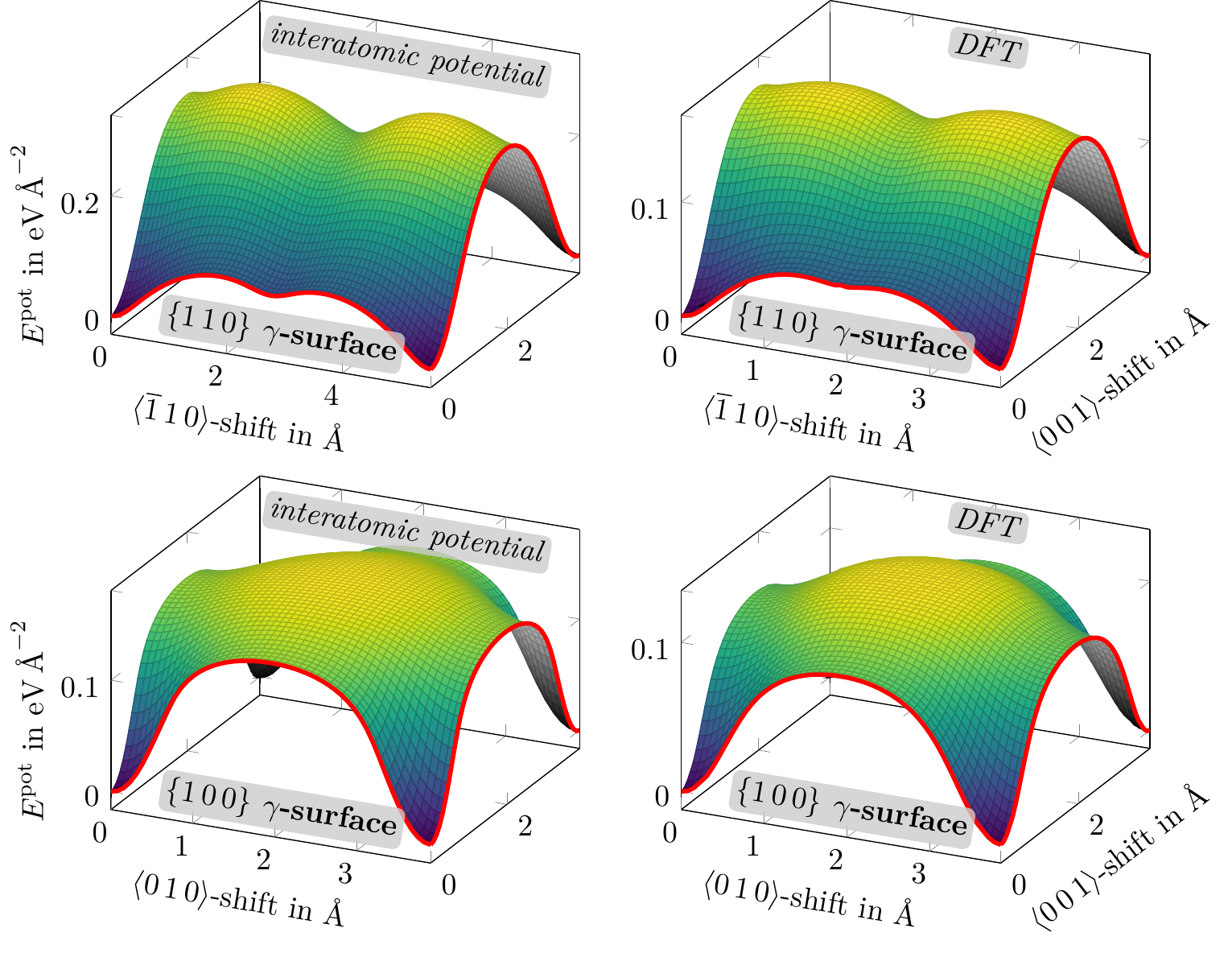}
            \caption{
                Stacking fault energy hypersurfaces calculated using the empirical pair potential by Thomas \emph{et al.} \cite{Thomas2005} in the left column.
                Analog results calculated using DFT with PBE GGA exchange correlation functional in the right column.
                Top: \hkl{110} surface with a distinct local minimum for shifts along the \hkl<110> direction.
                Bottom: \hkl{100} surface with shifts along \hkl<010> and \hkl<001> direction yield no metastable stacking faults.
                The general features are very well produced.
                Results are not converged with simulation cell size; see text for details.
            }
            \label{fig:sf_surface}
        \end{figure*}

        Using the interatomic potential by Thomas \emph{et al.} \cite{Thomas2005} and molecular statics methods the generalized stacking fault energy, called $\gamma$-surface, is computed for the \hkl(100) and \hkl(110) glide interfaces of SrTiO\textsubscript{3}, see \Cref{fig:sf_surface}.
        For comparison, calculations performed using the density-functional theory method are also displayed. 
        Both methods yield the same qualitative results which gives us confidence in the reliable modelling of dislocation structures using the classical potential.
        Moreover, the results are in line with the published literature \cite{Hirel2010}. The important feature reflected in both types of calculations is the local minimum present in the \hkl(110) $\gamma$-surface along the \hkl[1-10] direction.

        This metastable stacking fault energy indicates that the \hkl<110>\hkl{1-10} dislocation is able to split into two equal and collinear partial dislocations of half length \cite{Vitek1968}.
        In between the two partials a stacking fault will be created tieing them together.
        The splitting of the dislocation's Burgers vector can be described by the following pseudo-reaction:
        \begin{align}
            \hkl<110> \rightarrow \frac{1}{2}\hkl<110> + {\text{\textit{SF}}}_{\hkl{1-10}} + \frac{1}{2}\hkl<110> \text{.}
        \end{align}
        Assuming that the two partial dislocations of the \hkl<110>\hkl{1-10} system (types $\mathcal{A}$, $\mathcal{B}$ \& $\mathcal{C}$) are far  apart and do not interact, their respective energy per unit length can be formulated:
        \begin{alignat}{5}
            &\Gamma^{\text{full}} &&= &&\Gamma^{\text{full}}_{\text{elast}} &&+ &&\Gamma^{\text{full}}_{\text{core}} \text{ ,} \label{eq:line_energy_full}\\
            &\Gamma^{\text{split}} &&= 2 &&\Gamma^{\text{partial}}_{\text{elast}} &&+ 2 &&\Gamma^{\text{partial}}_{\text{core}} + d \gamma \text{ .} \label{eq:line_energy_partials}
        \end{alignat}
        Here the energy per dislocation line $\Gamma^i$ has been decomposed into an elastic part $\Gamma^i_{\text{elast}}$, a dislocation core contribution $\Gamma^i_{\text{core}}$ and the stacking fault energy $\gamma$ multiplied by the dislocation splitting distance $d$.
        Although we will see below that the partial dislocations are in fact interacting, we stay with this assumption for now.
        It is also fair to assume that the elastic energy is the largest contribution to the line energy \cite{Mitchell1985,Nabarro1997}.
        Along the (100) directions the $\gamma$-surface doesn't show a local minimum. Therefore,  dislocation with \hkl<001> Burgers vector on  \hkl{110} and \hkl{100} planes stay compact as  full dislocations. Also, dislocation with \hkl<110> Burgers remain full dislocations dislocations on a \hkl{001} plane.

        At this point we have consistently established that the dislocations with the lowest energy in SrTiO\textsubscript{3} are of \hkl<100>\hkl{001}, \hkl<001>\hkl{1-10} and split \hkl<110>\hkl{1-10} type, see \Cref{tab:squared_b}.
        Next, we consider a less obvious argument in order to explain the type of dislocations we expect to observe.
        Since dislocations carrying plasticity are introduced and moved through the crystal by mechanical force we make a rough estimate of their kinetic behavior.
        In doing so, we consider the \textquote{minimum stress to keep a dislocation moving} $\tau_m$ \cite{Huntington1955}.
        This so-called Peierls stress \cite{Nabarro1997, Hull2011} is estimated by
        \begin{align}
            \tau_m \propto \exp{\left(\frac{-2\pi h}{b}\right)}\text{,} \label{eq:Peierls_stress_proportionality}
        \end{align}
        with the length of the Burgers vector $b$ and distance $h$ between equivalent lattice planes.
        Normalized values of the Peierls stress $\tau_m / \tau_m^0$ are given in \Cref{tab:dislocation_Peierls_estimate} for the different dislocation types.
        Note, that the lattice spacing is $a/2$ when the glide plane is a \hkl{001} plane and $a/\sqrt{2}$ for \hkl{1-10} glide planes.
        
        \begin{table}[ht] 
            \caption{For the slip systems indicated in \Cref{tab:squared_b}, Burgers vector length $b$, distance between glide planes $h$, the proportionality factor of the Peierls stress, see Equation \eqref{eq:Peierls_stress_proportionality}, and the reduced Peierls stress $\tau_m/\tau_m^0$ are given. Here the $\tau_m$ has been normalized to the Peierls stress for the most simple dislocation type $\tau_m^0$ of the \hkl<100>\hkl{001} slip system. Dislocations with glide dissociation, see last line, can have strongly reduced Peierls barriers.}
            \label{tab:dislocation_Peierls_estimate}
            \begin{center}
                \begin{tabular}{r | l l l l}
                    \toprule
                    dislocation & $b$ & $h$ & $\exp{\left(\frac{-2\pi h}{b}\right)}$ & $\tau_m/\tau_m^0$ \\
                    \midrule
                    \hkl<100>\hkl{001}  & $a$ & $a/2$ & \num{0.043} & \num{1} \\
                    \hkl<110>\hkl{001}  & $\sqrt{2}a$ & $a/2$ & \num{0.108} & \num{2.5} \\
                    \hkl<001>\hkl{1-10} & $a$ & $a/\sqrt{2}$ & \num{0.012} & \num{0.27} \\
                    \hkl<110>\hkl{1-10} & $\sqrt{2}a$ & $a/\sqrt{2}$ & \num{0.043} & \num{1} \\
                    $\frac{1}{2} \hkl<110>\hkl{1-10}$ & $a/\sqrt{2}$ & $a/\sqrt{2}$ & \num{0.002} & \num{0.043} \\
                    \bottomrule
                \end{tabular}
            \end{center}
        \end{table}

        The split dislocations show the lowest Peierls stresses and are, therefore, the most relevant types for plastic deformation in SrTiO\textsubscript{3}.
        Indeed, this is exactly what is observed in multiple studies that have characterized dislocations in plastically deformed SrTiO\textsubscript{3}  \cite{Brunner2001, Gumbsch2001, Taeri2004,Porz2021}.

        Note that the consideration regarding the dislocation structure made here rely on static considerations and assume an elastically continuous medium.
        In this following we will study these dislocations in atomistic simulations and discuss their energetics and properties at low temperature. 
        
        It is known from experiment, however, that at high temperatures exceeding \qty{1000}{\kelvin} dislocations in SrTiO\textsubscript{3} are predominantly of \hkl<100>\hkl{001} type. The detailed investigation of these dislocations is outside the scope of this publication.

    \subsection{Type $\mathcal{A}$ dislocation}
    \label{subsec:type_A}

        This type of dislocation with $\vec{b} = a$\hkl<110>, $\vec{t} = $ \hkl<001> and glide plane $\vec{n} = $ \hkl{1-10} is one of the best investigated dislocation types in SrTiO\textsubscript{3}.
        It is intimately connected to type $\mathcal{B}$ and $\mathcal{C}$ dislocations that share the same glide plane.
        In literature \cite{Brunner2001, Gumbsch2001, Taeri2004} it is proposed that the  $\mathcal{A}$ type is mainly is responsible for the obverved ductility of SrTiO\textsubscript{3}.
        These observations were made primarily in single crystals compressed along a \hkl<100> axis in the temperature regime from \qty{110}{\kelvin} to \qty{1000}{\kelvin}.
        It was long thought to be the most prominent edge dislocation type in SrTiO\textsubscript{3} appearing naturally in its glide dissociated form.

        It should be noted, however, that the glide dissociated \hkl<110>\hkl{1-10} dislocation with pure edge character has rarely been observed in experiment \cite{Yang2011a}.
        Only the climb dissociated variant has been realized in bicrystal experiments that were specifically designed to observe \hkl<110>\hkl{1-10} dislocations with pure edge character \cite{Zhang2002}.
        Notwithstanding, this glide dissociated form was often related to the low flow stress of SrTiO\textsubscript{3}, see discussion above.
        Thus, we ask, what is the equilibrium configuration of this specific dislocation.

        Even the pure edge type \hkl<110>\hkl{1-10} dislocation seems to be a rare case.
        This was shown by continued work of Sigle \emph{et al.} \cite{Sigle2006}. They are suggesting that the nature of the $\vec{b} = \hkl<110>$ dislocations actually varies strongly with temperature.
        Dislocations found in TEM images after compression were identified to be of predominantly screw type (type $\mathcal{B}$) below and edge type (type $\mathcal{A}$) above room temperature.
        At room temperature neither of the two types dominates.
        It was, thus, proposed that pure full edge dislocations tend to climb as temperature is increased, then become sessile and lead to embrittlement.
        Therefore, they are visible in samples loaded at high but less visible at low temperatures, where they remain mobile enough to leave the crystal.
        On the contrary, pure screw dislocations (type $\mathcal{B}$, details see below) observed after low temperature loading experiments suggest a low mobility at low and a high mobility at high temperatures.

        Some doubts on the exact equilibrium structure of this dislocation type have also been raised by Yang \emph{et al.} who performed compression of \hkl<100> oriented SrTiO\textsubscript{3} single crystals up to \qty{19}{\percent} \cite{Yang2011}.
        In their work they claim to find dislocations with $\vec{b} = \hkl<110>$ being predominantly configured as glide dissociated edge partials. However, they also mention \hkl<111> line vectors, i.e. mixed dislocations. The exact ratio of the different dislocation configurations remains, however, unclear.
        We, therefore, propose that dislocation type $\mathcal{A}$ and type $\mathcal{B}$ can be seen as limiting cases of the true dislocation structure.
        In between there is another distinct dislocation line orientation leading to a mixed dislocation character which is discussed as type $\mathcal{C}$ dislocations later.

        The reasons for the glide dissociated form of the type $\mathcal{A}$ dislocation (and types $\mathcal{B}$ and $\mathcal{C}$, correspondingly) and its connection to ductility are discussed in the following, and we will also make a quantitative estimate of the splitting distance below.
        
        Glide dissociation into two equal partials on an \hkl{1-10} plane can happen spontaneously for \hkl<110> edge (type $\mathcal{A}$) and screw dislocations (type $\mathcal{B}$) \cite{Hirel2010}.
        In order to explain the loss of ductility in SrTiO\textsubscript{3} at high temperatures it is proposed that the \hkl<110>\hkl{1-10} edge dislocations climb dissociate \cite{Taeri2004}.
        However, climb dissociation is subject to diffusion and thus only possible at elevated temperature.
        This has been observed using MD simulations at extremely high temperatures (and presumably extremely high pressures) which led to a constriction of the glide dissociated configuration and ultimately resulted in a climb dissociated configuration \cite{Hirel2016}.
        Nudged-elastic band calculations, additionally, suggest that the climb dissociated configuration is in fact energetically more favorable.
        The obvious question why the glide configuration forms in the first place is, however, not answered.

        To sum up, it remains unclear what the exact structure of this mechanically induced dislocation type under physically relevant conditions is: are they full dislocations or do they appear as partials in glide or climb dissociated form?
        And do they alter their nature spontaneously or react at elevated temperatures leading to decreased effective mobility and, thus, material embrittlement?

\begin{figure*}[htb] 
            \centering
            \includegraphics[]{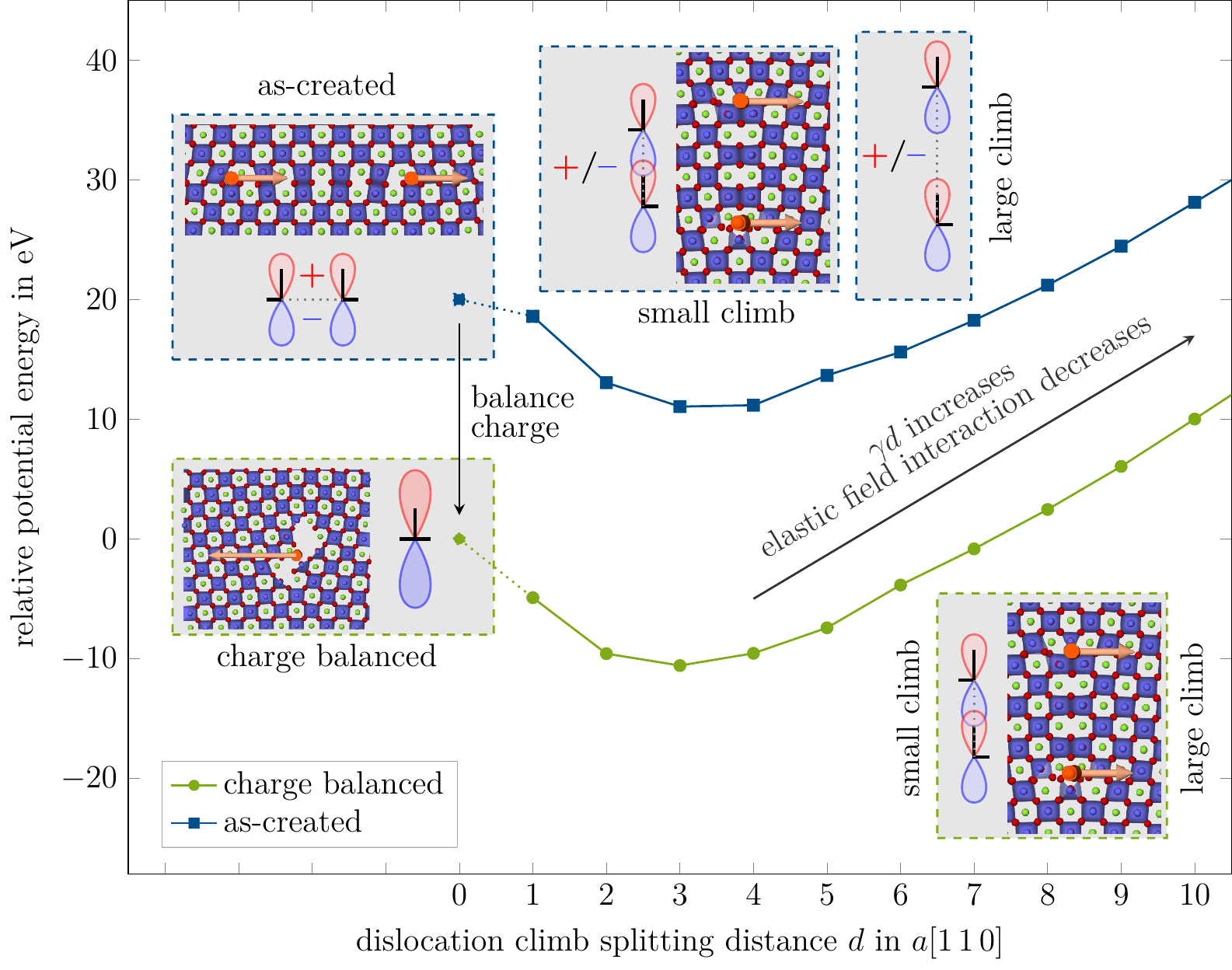}
            \caption{
                Comparison of energies of type $\mathcal{A}$ dislocations with and without charge balancing for different climb splitting distances.
                Each point represents the average of at least three independent samples.
                By balancing the charge on the dislocations the energy can be decreased.
                For the native dislocation the zero splitting point shows glide dissociation.
                An equilibrium climb splitting of three unit cells is evident.
                A growing stacking fault and decreasing elastic interaction increases energy at high climb splitting distances.
                }
            \label{fig:type_A_energies}
        \end{figure*}

        \paragraph{\textbf{Glide vs. climb}}
        A schematic representation of the full, glide dissociated, and climb dissociated configurations is given  in \Cref{fig:type_A_energies} together with atomistic structure models.
        The climb dissociated configuration is the result of a full edge dislocation or a glide dissociated pair of edge dislocation partials that climb dissociated at elevated temperatures.
        
        For glide dissociation the stacking fault between the two partials is on the glide plane and partial dislocations can either move one by one (as suggested by Ref. \cite{Hirel2012}) or in a correlated fashion (as suggested by Ref. \cite{Nabarro1997}).

        Also, in a climb dissociated form the stacking fault is on a \hkl{110} plane.
        The stacking fault is of the same crystallography as in the glide dissociated case.
        But now it is terminated by partial dislocations which have a glide plane different from the stacking fault plane.
        Motion of the dislocation in a conservative manner cannot occur since then the stacking fault had to be dragged through the crystal which, in turn, requires diffusion.
        The dislocation mobility is lost for this climb dissociated form and the material reacts macroscopically brittle to compression if not sufficient dislocation nucleation or activation of other dislocations can occur.

        Let us first address the glide dissociation from an analytical point of view.
        In the discussion above we have already noted that the Peierls barrier for a glide dissociated dislocation is lowered significantly compared to the full dislocation (see Equation \eqref{eq:Peierls_stress_proportionality} and \Cref{tab:dislocation_Peierls_estimate}).
        Regarding the excess energy due to dislocation dissociation we formulated Equations \eqref{eq:line_energy_full} and \eqref{eq:line_energy_partials}.
        These equations indicated that we expect the glide dissociation based on energy criteria alone.
        Dislocation splitting will occur if 
        \begin{align}
            \Gamma^{\text{split}} \leq \Gamma^{\text{full}} \text{,}
        \end{align}
        and if we assume that $\Gamma^{\text{full}}_{\text{core}} \approx 2 \Gamma^{\text{partial}}_{\text{core}} \ll \Gamma^i_{\text{elast}}$ we can rearrange to:
        \begin{align}
            0 \approx \Gamma^{\text{full}}_{\text{elast}} - \left( 2 \Gamma^{\text{partial}}_{\text{elast}} + d \gamma \right) \text{.}
        \end{align}
        Making use of the solution for the elastic line energy of the dislocation (cf. Ref. \cite[p.~71]{Anderson2017})) we arrive at
        \begin{align}
            \begin{split}
                0 =& \frac{\mu}{4 \pi \left(1-\nu\right)} \ln{\left( \frac{R}{r_{\text{full}}} \right)} b_{\text{full}}^2 \\
                &- \left[ 2 \frac{\mu}{4 \pi \left(1-\nu\right)}  \ln{\left( \frac{R}{r_{\text{partial}}} \right)} b_{\text{partial}}^2 + d \gamma \right] \text{,}
            \end{split}  
        \end{align}        
        which is then solved for $d = d_{\text{max}}$:
        \begin{align} \label{eq:splitting_distance_A_energy}
            \begin{split}
                d_{\text{max}} =& \frac{\mu}{4 \pi \left(1-\nu\right) \gamma} \left[ \ln{\left( \frac{R}{r_{\text{full}}} \right)} b_{\text{full}}^2 - 2 \ln{\left( \frac{R}{r_{\text{partial}}} \right)} b_{\text{partial}}^2 \right] \text{.}
            \end{split}
        \end{align}
        We use the values $b_{\text{full}} = \qty{3.905}{\angstrom}$, $b_{\text{partial}} = b_{\text{full}}/2$, $\mu = \qty{107}{\giga\pascal}$, $\nu = 0.26$, and $\gamma = \qty{1.225}{\joule\per\square\meter}$ for the given interatomic potential, insert $R \approx \qty{275.4}{\angstrom}$ which is half the distance between the dislocations and assume $r_{\text{partial}} = b_{\text{partial}}$ and $r_{\text{full}} = b_{\text{full}}$.
        It results in a value of $d_{\text{max}} = \qty{46.1}{\angstrom} \approx \qty{11}{} - \qty{12}{unit\, cells}$ as an upper limit for the partial dislocation separation distance.
        According to Hull and Bacon this is a good approximation for the equilibrium splitting distance \cite{Hull2011}.

        However, we can probably do much better by not considering the energy balance but the force equilibrium.
        The force $f_2$ that is exerted on the second partial in a split configuration is given by
        \begin{align}
            f_2 = -\gamma + f^{\text{PK}}_2 \text{,}
        \end{align}
        where $\gamma$ is the stacking fault energy that ties the partial dislocations together and $f^{\text{PK}}$ is the Peach-Koehler force driving the partials apart.
        In equilibrium the force $f_2$ needs to vanish.
        Making use of the Peach-Koehler force $f^{\text{PK}}_2 = \left( \vec{b}_2 \cdot \doubleunderline{\sigma_1} \right) \times \xi_2$ a solution for the interaction force between the partial dislocations is derived (cf. Ref. \cite[p.~112]{Anderson2017}), which reduces to:
        \begin{align}
            d = \frac{a^2 \mu}{4 \pi \left(1-\nu\right) \gamma} \text{.} \label{eq:splitting_distance_A_force}
        \end{align}
        With the values from above we arrive at an expected splitting distance of $d = \qty{14.3}{\angstrom} \approx \qty{3}{} - \qty{4}{unit\, cells}$.
        Apparently, the two estimates give quite varying results.
        Consequently, we will try to obtain more reliable values by the atomistic simulations.

        Also, for the case of climb dissociation of the type $\mathcal{A}$ dislocation the atomistic simulations will be helpful.
        As we can see in \Cref{fig:type_A_energies} the glide and climb stacking faults are of the exact same crystallography and, therefore, have the exact same energy.
        But, as indicated in \Cref{fig:type_A_energies} by red and blue regions, the areas around the partial dislocations, where compressive and tensile stress occur, partly overlap.
        The stress and strain fields are assumed to be linearly additive, resulting in decreased stress and strain below the top partial and above the bottom partial (\Cref{fig:type_A_energies} "small climb") compared to the glide dissociated configuration (\Cref{fig:type_A_energies} "as-created").
        Elastic energy is thereby reduced.
        Predictions about the splitting distance in this case are not trivial though.
        Consider the limiting cases of dislocation splitting, i.e., infinitely small and infinitely far splitting.
        At infinitely small splitting we end up with a full dislocation and loose the advantage of split dislocations decreasing elastic energy.
        At infinitely large splitting we loose the advantage of overlapping and cancelling elastic fields.
        This type of splitting will, thus, be the second point of concern for the simulations below.

\begin{figure}[htb] 
            \centering
            \includegraphics[]{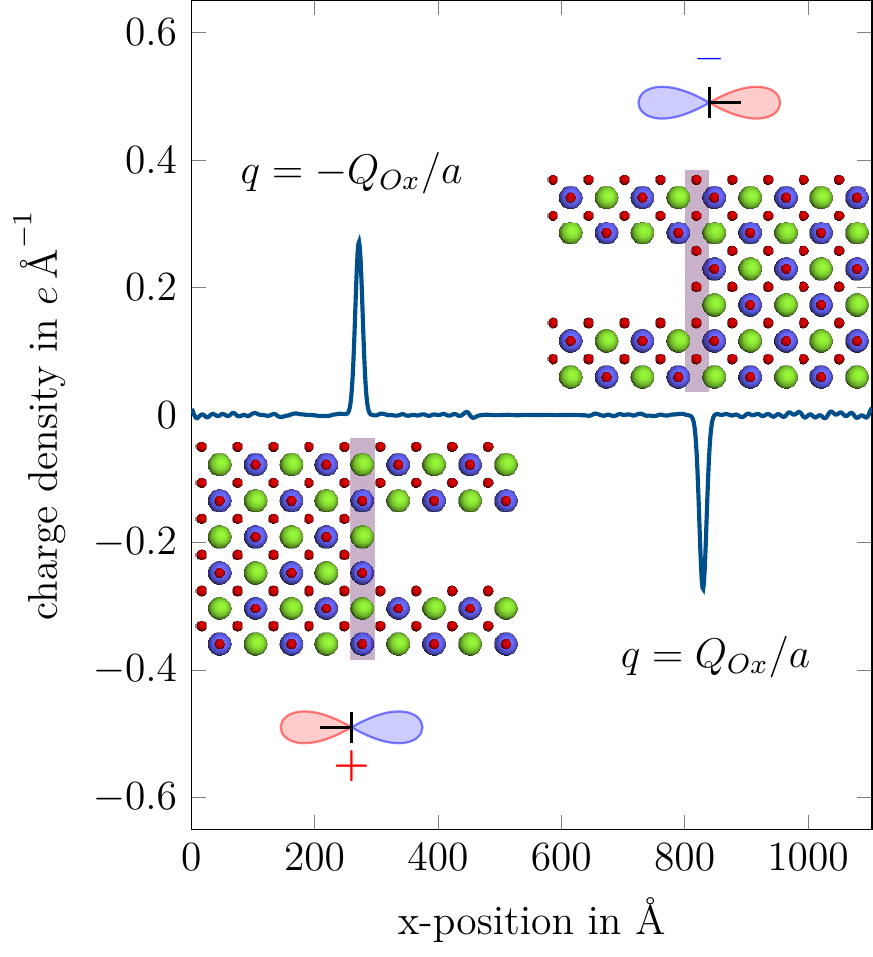}        
            \caption{
                Charge density in a simulation cell containing a type $\mathcal{A}$ dislocation dipole.
                In order to obtain a smooth charge density that can be binned along the x-direction, the ionic point charges have been smeared out be a Gaussian distribution.
                Additionally, the integrated charges $q$ for each dislocation and a representation of the terminal layer of the dislocation cut are given.}
            \label{fig:type_A_charge_imbalance}
        \end{figure}

        \paragraph{\textbf{Simulation}}
        The dislocation is created according to the above procedure of removing atoms and equilibrated.
        In \Cref{fig:type_A_charge_imbalance}, there is an excerpt from the simulation cell with a full unit cell layer removed in the as-created structure.
        The atoms relax into the gap, form a dislocation and spontaneously glide dissociate.
        Thereby as-created full dislocations spontaneously split into two glide dissociated partials which reduces the elastic energy contribution.
        The full dislocation is not stable, as suggested by the considerations of elastic energy in \Cref{fig:type_A_energies}.
        We also note that the glide splitting distance measured here (\(\qty{15.6}{} - \qty{20.5}{\angstrom}\)) is in good agreement with the estimations made on the basis of the forces exerted by the partial dislocations on each other.

        At the same time this reveals that the climb dissociated configuration does not spontaneously form. 
        In order to avoid high temperatures and pressures to enable the climb dissociation (see e.g. Ref. \cite{Hirel2016}) we prepare the simulation cell with a priori climb dissociated dislocations.
        To construct climb dissociated dislocations half of the atoms on the full cut plane, i.e., a layer for height $\frac{1}{2}\hkl<110>$, is shifted in increments of unit cells relative to the other half of the removed atoms.
        In \Cref{fig:type_A_charge_imbalance} this corresponds to shifting half of the plane that has been cut to either the left or the right side incrementally (analog to \Cref{fig:type_E_variants} right).
        When these structures are relaxed, the climb dissociated configuration forms.

        Now, to compare the stability of the glide dissociated structure and the climb dissociated structure we compare the energies of the simulation cell in \Cref{fig:type_A_energies}.

        These results clearly confirm our hypothesis: the climb dissociated dislocation is lower in energy than the glide dissociated dislocation and there is an optimum climb dissociation distance of about $\qty{3}{} - \qty{4}{unit\, cells}$.
        At lower splitting distances the elastic fields overlap constructively and raise the elastic energy of the system.
        At larger splitting distances the stacking fault length increases, i.e. $d\gamma$ increases, and the destructive overlap of the partial dislocations' elastic fields also decreases.
        The glide dissociated dislocation that is made responsible for the ductility of SrTiO\textsubscript{3} is, therefore, only meta-stable.
        When temperature increases to provide enough activation it is probable that type $\mathcal{A}$ dislocations constrict and then split into the energetically favorable climb dissociated configuration \cite{Hirel2016}.
        These dislocations are then sessile.

        \paragraph{\textbf{Charge at dislocation}}
        A further complication with this dislocation structure arises due to the ionic nature of the solid: there are different options to apply the cut plane to introduce the dislocation in the simulation cell.
        With a closer look at \Cref{fig:type_A_charge_imbalance} top we recognize that the terminal plane of the cut can either be terminated by a cation rich layer (right cut) or by an anion rich layer (left cut).
        The as-created dislocation are, therefore, not charge balanced.
        
        We observe different glide dissociation splitting distances at the positive end (\qty{5}{unit\, cells}) vs. the negative end (\qty{4}{unit\, cells}). 
        Apparently the charge state has some influence on the dislocation structure.
        Note that it may also have a direct influence on the dislocation mobility, as discussed by Eshelby \emph{et al.} \cite{Eshelby1958}.

        A charge imbalance is the reason for a force driving ion flux.
        It is, therefore, reasonable to assume that dislocations in static equilibrium will be charge compensated. 
        In the simulation the charge imbalance can easily be compensated by (manually) moving one oxygen ion per unit cell dislocation line length from the layer of oxygen ions at the negative end to the positive end prior to dislocations equilibration (cf. Ref. \cite{Marrocchelli2015}).
        It was additionally confirmed that the balancing of charges leads to minimum energy because of reduced electrostatic energy.

        Interestingly, the charge balanced dislocations do not show spontaneous glide dissociation.
        The overall energy is still decreased by climb dissociation compared to the as-created full dislocation.
        The energies for the as-created as well as the charge balanced dislocation with and without climb dissociation are shown in \Cref{fig:type_A_energies}.

        The implications from this finding are far-reaching.
        For now assume that glide-dissociated dislocations of type $\mathcal{A}$ are the dislocation type that usually governs ductility in SrTiO\textsubscript{3}.
        When the dislocations are introduced during deformation they move through the crystal as glide dissociated partials \cite{Porz2021}.
        Their quick motion does not allow them to find an equilibrium configuration, and they possess excess charge that stabilizes the glide dissociation.
        After deformation the sample is then allowed to equilibrate, i.e., oxygen ions diffuse through the lattice and change the dislocation core structure.
        Thereby the dislocation would constrict to a full dislocation while lowering its energy at the same time (see "balance charge" arrow in \Cref{fig:type_A_energies}).
        When such a constricted dislocation is supposed to contribute to further deformation it first would have to dissociate into two partials again lowering its Peierls barrier and then continue movement in its glide dissociated form.

        This gives a hint for the ductile to brittle transition in SrTiO\textsubscript{3}.
        In a first step constriction of dissociated dislocations raises the Peierls barrier which is facilitated by oxygen diffusion.
        In a successive step diffusion of all species (Sr, Ti and O) at elevated temperatures enables climb dissociation which makes the dislocations sessile.
        This tendency to climb is clearly evident from our data and, therefore, supports earlier studies that have suspected such a mechanism \cite{Gumbsch2001,Taeri2004,Hirel2016}.
        Note that other defect types such as dopants have not been considered so far and possibly strongly alter microscopic and macroscopic behavior.

        The findings by Porz \emph{et al.} are consistent with this observation and add another aspect to dislocation based plasticity \cite{Porz2021}.
        Climbing of dislocation partials during crystal deformation and dislocation motion has been identified as a means of creating new dislocation that enable macroscopic deformation.
        This climb motion can now be clearly related to the instability of the glide dissociated type $\mathcal{A}$ dislocation.

        \paragraph{\textbf{Loading the dislocation}}
        The three dislocation configurations, i.e., as-created glide dissociated, charge balanced full, and charge balanced with optimum climb dissociation, have been loaded by a shear stress $\tau_{xy}$ at low temperature.
        As expected, the sample with the charge balanced full dislocation only ruptures around \qty{6}{\giga\pascal} without dislocation motion.
        In the case of the climb dissociated dislocation they act as nucleation centers for mobile glide dissociated dislocations that get emitted at a load of \qty{4.5}{\giga\pascal}.
        This stress level is likely to be only relevant in nanoindentation experiments or when stress fields of external applied stress and other dislocations overlap \cite{Porz2021}.
        
        Only for the glide dissociated configuration the dislocation starts moving between \qty{1.20}{\giga\pascal} and \qty{1.25}{\giga\pascal} which is in line with the static result of \qty{1.5}{\giga\pascal} from literature \cite{Hirel2012}.
        However, during its motion the dislocation does emit oxygen vacancies that are left behind in the dislocation's track.
        Already after travelling approx. \qty{25}{\nano\meter} the dislocation has equilibrated its charge by emission of oxygen vacancies.
        Despite still being in the glide dissociated configuration, the dislocation motion stops at this point.
        Therefore, the dislocation's mobility is indeed lowered by charge balancing.

\begin{figure}[htb] 
            \centering
            \includegraphics[]{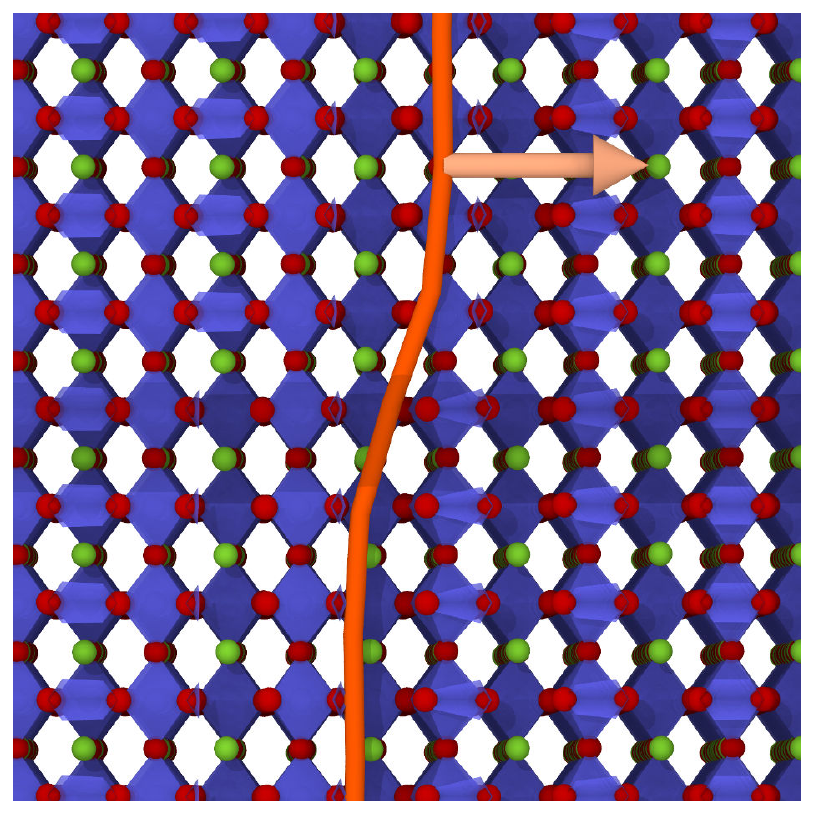}
            \caption{Motion of a type $\mathcal{A}$ dislocation by kinking.}
            \label{fig:type_A_kink}
        \end{figure}

        In the large simulation cell, where the dislocation line length was increased from 2~unit~cells to 60~unit~cells, the stress to move a dislocation is slightly reduced to \qty{1.05}{\giga\pascal} -- \qty{1.10}{\giga\pascal}.
        Here the dislocation motion is clearly proceeding by nucleation of kinks and subsequent kink motion as shown in \Cref{fig:type_A_kink}.
        Nevertheless, the dislocation lines remain rather straight and kinks of double height are virtually absent.
        From this we conclude that the kink formation is the limiting step of this dislocation motion.

        Performing the same experiments at \qty{1500}{\kelvin} reduces the $\tau^m_{xy}$ to the range of \qty{120}{\mega\pascal} to \qty{150}{\mega\pascal}.
        Especially for the large dislocation line length, the effect of immobilization by emission of point defects is even more pronounced.
        Oxygen vacancies are emitted at a much higher rate than at low temperatures.
        As a result, the charge equilibrated dislocations already associate to form a full dislocation after travelling a few nanometers and become immobile.

        \paragraph{\textbf{Outlook}}
        Based on our results, we suggest that the importance of glide dissociated type \(\mathcal{A}\) dislocations has been overestimated in literature.
        Rather note that experimental evidence for climb dissociated dislocations of this type seems to be available \cite{Taeri2004}.
        In order to explain this configuration Jin \emph{et al.} have given an alternative explanation that has to do with type $\mathcal{C}$ dislocations and will be treated below \cite{Jin2013}.
        Since this explanation avoids the meta-stable glide dissociated type $\mathcal{A}$ dislocations, it is deemed more probable.

    \subsection{Type $\mathcal{B}$ dislocation}

        Type $\mathcal{B}$ dislocations are the screw analog to type $\mathcal{A}$ dislocations, i.e., the $\vec{b} = a$\hkl<110> Burgers vector is aligned with the $\vec{t} = $ \hkl<110> line vector and the preferred glide plane is also $\vec{n} = $ \hkl{1-10}, see \Cref{fig:types_overview}.

        In the framework of dislocation motion by dislocation kinking, type $\mathcal{A}$ (edge) dislocations and type $\mathcal{B}$ (screw) dislocations are naturally interchangeable \cite{Sigle2006}.
        The dominance of either type can, thus, be attributed to the likelihood of kink formation vs. kink migration \cite{Brunner2008,Yang2011a}.
        The single pure type $\mathcal{B}$ screw may, thus, be viewed as another limiting case to the true dislocation structure.
        In the following section the relation to the intermediate type $\mathcal{C}$ dislocations will also be discussed.

        In further studies hints have been found that the predominance of either type $\mathcal{A}$ or type $\mathcal{B}$ dislocations is temperature dependent \cite{Sigle2006,Castillo-Rodriguez2010,Castillo-Rodriguez2011}.
        These Refs. find type $\mathcal{B}$ to be dominant in experiment after sub-room temperature deformation.
        According to their findings it dissociates on a \hkl{1-10} plane with an equilibrium splitting distance around \qty{4.2}{\nano\meter}.
        They confirm that pairs of partials, if having opposite Burgers vector, preferentially arrange in a dipolar manner with dipole heights of several nanometers at room temperature.

        Hirel \emph{et al.} have performed investigations employing MD simulations \cite{Hirel2012}.
        In this work it is found that type $\mathcal{B}$ dislocations spontaneously glide dissociate on a \hkl{1-10} plane with dissociation distances about \qty{25}{\percent} less than the type $\mathcal{A}$ edge dislocations.
        This difference in splitting distance comes close to the expected values from our analytical estimates.
        Referring to Equations~\eqref{eq:splitting_distance_A_energy} and \eqref{eq:splitting_distance_A_force} that we used to estimate the splitting distance of type $\mathcal{A}$ dislocations we note that they need to be modified by omitting the factor $(1-\nu)$ to account for the case of screw dislocations \cite[pp.~59~\&~110]{Anderson2017}. With $\nu = 0.26$ a reduction of splitting distance of approximately \qty{26}{\percent} is expected:

        \begin{align}
            d = \frac{a^2 \mu}{4 \pi \gamma} \text{.} \label{eq:splitting_distance_B_force}
        \end{align}
        With the values from above we arrive at an expected splitting distance of $d = \qty{10.8}{\angstrom} \approx \qty{2}{} - \qty{3}{unit\, cells}$.

        At the same time Hirel \emph{et al.} estimate the dislocation mobilities of type $\mathcal{A}$ and $\mathcal{B}$ dislocations and find the latter to be less mobile at low temperature.
        This observation is in line with the experimental finding in Ref. \cite{Castillo-Rodriguez2010} because the less mobile dislocation type is more likely to be observed in TEM images after deformation.

        A further hint to the importance of this dislocation type can also be gained from nanoindentation experiments.
        In an in situ TEM nanoindentation experiment it was shown that -- for indentation on a \hkl{001} surface -- deformation occurs by dislocations with $\vec{b} = \hkl<110>$ moving on \hkl{1-10} planes that are of \textquote{near-screw type} \cite{Kondo2012}. The line direction is, however, strongly influenced by surface effects in this experiment which tend to favor the screw component.

        \paragraph{\textbf{Simulation}}
        In our MD simulation the screw dislocation can be introduced in a perfect crystal by applying the analytical displacement field of a full dislocation to the atomic positions in the crystal.
        Note that screw dislocation cores differ fundamentally from their edge counterpart.
        They are always charge balanced independent of where the cut is made.
        However, there is no climb dissociation possible for type \(\mathcal{B}\).

        \begin{figure}[htb]
            \centering
            \includegraphics[]{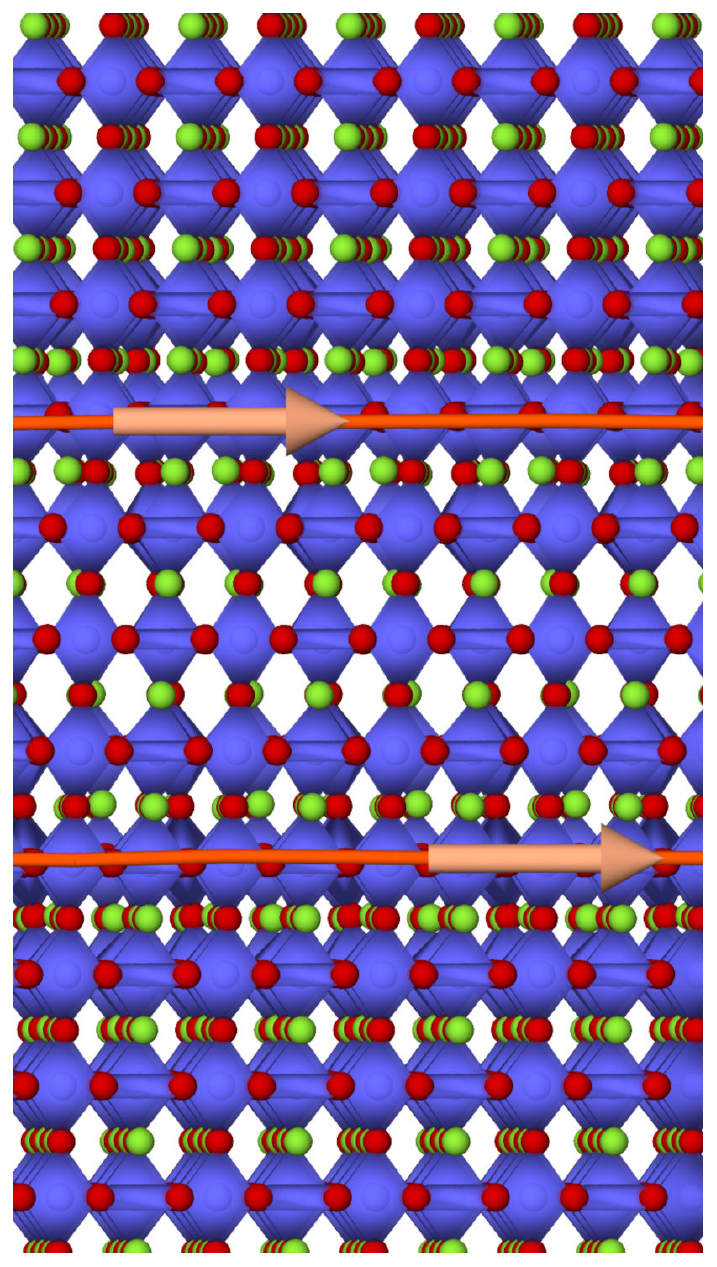}       
            \caption{
                Type $\mathcal{B}$ dislocation in its glide dissociated form cut from a larger sample.
                The oxygen octahedra, dislocation lines and Burgers vectors are added as a guide to the eye.
                }
            \label{fig:type_B_perspectives}
        \end{figure}

        In our own calculations we could in principle confirm the findings from literature:
        Type $\mathcal{B}$ dislocations dissociate on a \hkl{1-10} plane into two equal and collinear partials separated by a stacking fault.
        The splitting distance is approximately equal to the edge dislocation with a value of approx. \qty{19}{\angstrom}.
        In fact, from linear elasticity a smaller value was expected.
        Yet, since type $\mathcal{B}$ dislocations do not carry charge, they are difficult to compare to type  $\mathcal{A}$: Glide dissociation is always expected to occur, unless impurities, dopants or non-stoichiometry in SrTiO\textsubscript{3} alter the stacking fault energy.
        The implications of dislocation glide dissociation for the reduction of the Peierls barrier as discussed above still hold true for this dislocation type \cite{Hull2011}.
        Because type $\mathcal{B}$ dislocations are not prone to re-associating into a full dislocation and do not exhibit climb, they are expected to stay mobile even after temperature treatments that would immobilize type $\mathcal{A}$ dislocations.

        \paragraph{\textbf{Loading the dislocation}}
        For the loading of this dislocation type with $\tau_{yz}$ we find a Peierls stress of \qty{2.50}{\giga\pascal} to \qty{2.75}{\giga\pascal} in the low temperature simulations independent of the system size, i.e., the dislocation line length.
        This also in line with Ref. \cite{Hirel2012} who find a value of \qty{2.64}{\giga\pascal} at \qty{0}{\kelvin}.
        In the high temperature regime this reduces to a $\tau^m_{yz}$ of about \qty{120}{\mega\pascal}.
        Even in the large samples the dislocations remain straight and kinks of double height never form.
        We, therefore, conclude that kink nucleation is the limiting step.

    \subsection{Type $\mathcal{C}$ dislocation} 
       
        So far we have looked at two dislocation types that feature a \hkl<110> Burgers vector and \hkl{1-10} glide plane with pure edge character (type $\mathcal{A}$) and pure screw character (type $\mathcal{B}$).
        In between these limiting cases there is at least one other prominent type of dislocation belonging to the same family, see \Cref{fig:types_overview}.
        This one has a \hkl<111> line vector and is, thus, of mixed character \cite{Jin2013}.
        The Burgers vector can be decomposed into its edge and screw contributions by projection onto the line vector according to the following equations:
        \begin{alignat}{3}
            &\vec{b}     &&= \vec{b}_{\text{edge}} &&+ \vec{b}_{\text{screw}} \text{ ,} \\
            &a \hkl<110> &&= \frac{a}{3} \hkl<11-2> &&+ \frac{2a}{3} \hkl<111> \text{.}
        \end{alignat}

        It has been first investigated in detail by Jin \emph{et al.} in Ref. \cite{Jin2013} but closer examination of prior publications also reveals hints to this type of dislocation \cite{Brunner2001, Gumbsch2001, Taeri2004, Yang2011}.
        Yet the nature of this dislocation type and possible connections to other dislocation types has remained unclear for some time.

        In their thorough TEM study Jin \emph{et al.} present evidence, that, indeed, the primary slip system is the \hkl<110>\hkl{1-10} system.
        But the preferential line direction is the \hkl<111> direction, making the dislocation have a mixed character.
        Consequently, many of the above made conclusions and results need to be revisited when it comes to plasticity.
        It is also presented how edge dislocation dipoles in a seemingly climb dissociated configuration are the consequence of a reaction of two mixed dislocations, how mixed dislocations on parallel slip planes lead to favorable dislocation stacking, and how anti-parallel dislocations form dipolar arrangements -- all keeping in mind the \hkl<111> line direction.

        In the context of dislocation line orientation also nanoindentation experiments have been conducted by Javaid \emph{et al.} \cite{Javaid2017}.
        Their molecular dynamics simulations on the dislocations in the bulk suggest that dislocations below the surface of \hkl<110>\hkl{1-10} character oriented preferentially along \hkl<111> line directions and dislocations do not split into partials.
        Based on the given evidence we find great interest in modelling this type $\mathcal{C}$ dislocation explicitly.

        \begin{figure*} 
            \centering
            \includegraphics[]{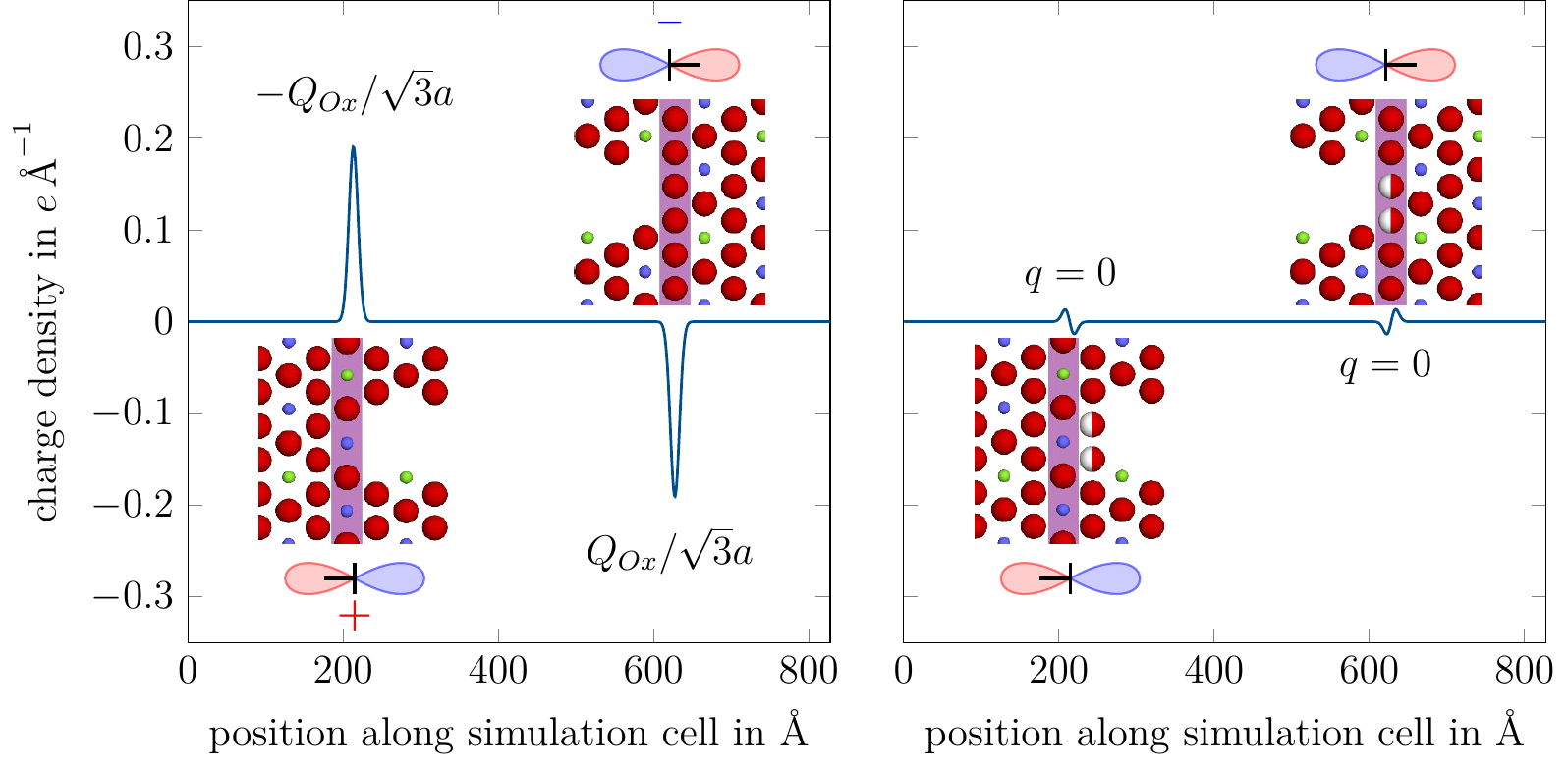}        
            \caption{Charge density in a simulation cell containing a type $\mathcal{C}$ dislocation dipole. In order to obtain a smooth charge density that can be binned along the x-direction, the ionic point charges have been smeared out be a Gaussian distribution. Additionally, the integrated charges $q$ for each dislocation and a representation of the terminal layer of the dislocation cut are given. Left: charge at the dislocation core is high in the as-created dislocation. Center: when half an oxygen column is removed from the negatively charged core to the positively charged core the dislocation core charge is reduced by half. Right: the dislocation cores are charge balanced when the oxygen ions of the terminal plane are equally distributed among the two dislocations.}
            \label{fig:type_C_charge_imbalance_variants}
        \end{figure*}

        \paragraph{\textbf{Simulation}}
        We created this dislocation again in an appropriate slab by deleting atoms for the edge component and application of the analytical displacement field for the screw component.
        Dislocation dipoles created in such a manner are charged, due to the edge contribution to the Burgers vector.
        This behavior is analog to the type $\mathcal{A}$ dislocations analyzed above.
        Thus, in order to balance charge individual oxygen atoms are shifted from one dislocation to the other, such that the terminal oxygen plane at one cut surface is only half occupied, see \Cref{fig:type_C_charge_imbalance_variants}.
        Tests are performed regarding the dislocation behavior without charge balance (zero oxygen ions shifted, left in \Cref{fig:type_C_charge_imbalance_variants}) and with charge balance (half of the oxygen columns shifted, right in \Cref{fig:type_C_charge_imbalance_variants}).
        
\begin{figure}[htb]
            \centering
            \includegraphics[]{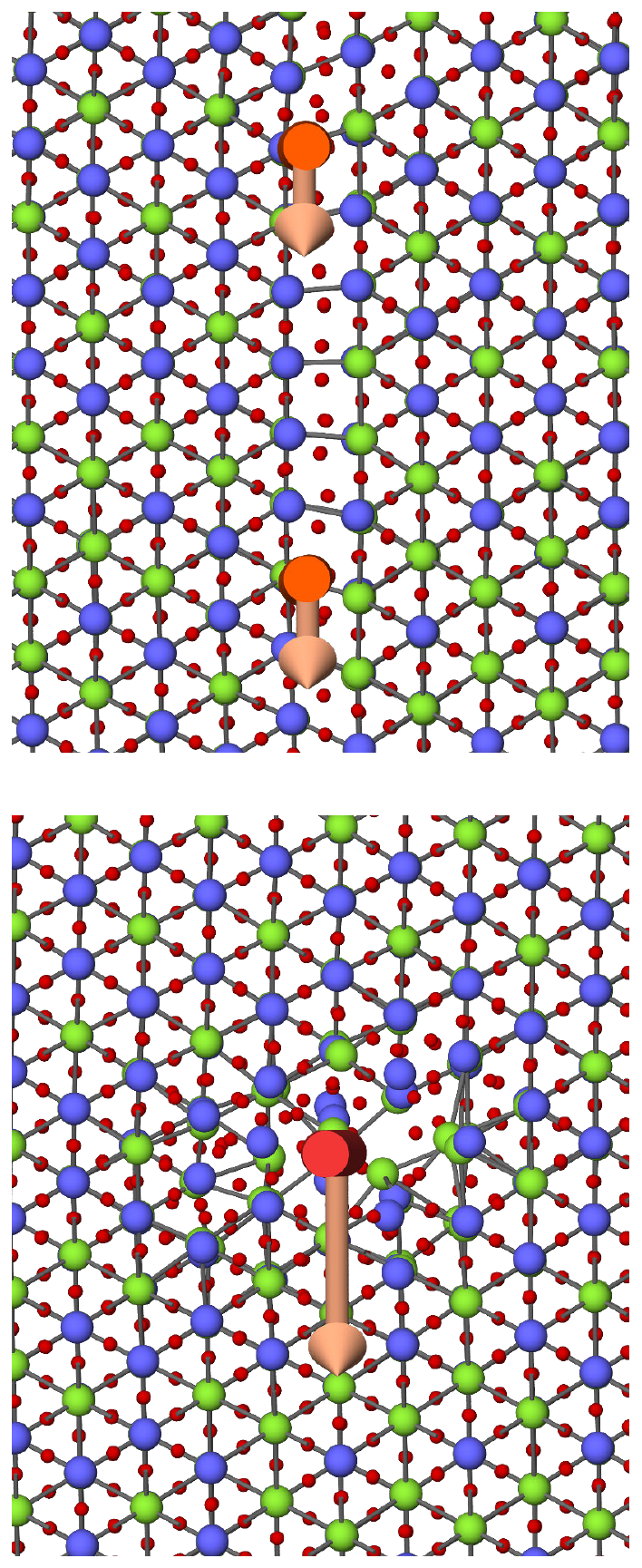}
            \caption[]{Top: glide dissociated type $\mathcal{C}$ dislocation after low temperature quasi-static relaxation. Bottom: the same dislocation after long high temperature annealing where the dislocation dipole has constricted to form a full dislocation with less ordered core structure. Dislocation lines and Burgers vectors are indicated.}
            \label{fig:type_C_annealing}
        \end{figure}

        Dislocation with and without charge both have their minimum energy configuration in a glide dissociated state.
        Splitting occurs as above into two equal collinear partials with $\frac{1}{2} \hkl<110>$ Burgers vector on a \hkl{1-10} plane by creating a stacking fault on this plane, \Cref{fig:type_C_annealing} top.
        The splitting distance is about \qty{19}{\angstrom} or \qty{4}{} -- \qty{5}{unit\,cells}, which is comparable to the \qty{25}{\angstrom} reported by Jin \emph{et al.} \cite{Jin2013}.
        In their publication they report an estimated $\gamma = \qty{0.606 \pm 0.077}{\joule\per\square\meter}$ which is about twice as high as featured by the interatomic potential.
        The lower splitting distance in the simulation confirms this, see \Cref{eq:splitting_distance_A_force}.
        Note that the experimental stacking fault energy is well comparable to our rough DFT-based estimate of \qty{0.795}{\joule\per\square\meter}, see \Cref{fig:sf_surface}.

        The behavior of type $\mathcal{C}$ is similar to type $\mathcal{A}$ dislocations which also tend to split on a \hkl{110} plane.
        And just like the type $\mathcal{A}$ dislocations, the shift on the \hkl{110} plane is along a \hkl<1-10> direction, corresponding to a local minimum on the stacking fault energy hypersurface, see \Cref{fig:sf_surface}.
        The stacking faults in type $\mathcal{A}$ and type $\mathcal{C}$ glide dissociation are, therefore, identical.

        Although the optimum configuration is the glide dissociated configuration irrespective of the charge state, the total energy does depend on the dislocation charge.
        First, the optimum energy of the charge balanced configuration is lower owing to the absence of an electrostatic dipole between the positively and the negatively charged dislocation.
        Second, compared to the full dislocation, dislocation splitting reduces energy by \qty{0.52}{\electronvolt\per\angstrom} in the uncharged and by \qty{1.18}{\electronvolt\per\angstrom} in the charged case, measured as potential energy per dislocation line length for our simulation cell size.
        Thus, charge balancing reduces the stability of the glide dissociated configuration compared to the full dislocation.

        Since the energy difference from glide dissociated to full dislocation is small, it may be overcome at finite temperatures, leading to a full dislocation.
        Full dislocations of this type are expected to have much higher Peierls barriers and, therefore, inhibit plastic deformation.
        Consequently, we perform annealing experiments analog to Ref. \cite{Hirel2016}.
        Samples with different charge compensations are are heated to \qty{2500}{\kelvin} and annealed for at least \qty{600}{\pico\second}.
        During this time almost all the dislocations constrict, i.e., the partial dislocations associate to form a full dislocation with a disordered core structure, see \Cref{fig:type_C_annealing}.
        We confirmed that even after the core rearrangement at high temperature, the full dislocation in any charge state is still energetically unfavorable compared the glide dissociated variant.
        Therefore, we suspect that the full dislocation core is only stabilized by entropy at finite temperatures.

        Climbing of dislocations or climb dissociation did not occur during the annealing.
        It was, nevertheless, observed that in case of dislocations with strong charge imbalance oxygen vacancy diffusion happens in order to lower the charge of the positively charged dislocation cores, also see discussion on dislocation type $\mathcal{D}$ below.
        The full dislocation core is, thus, stabilized by attaining a favorable charge state as well as possible entropic effects relevant at the comparably disordered dislocation core structures.
        In essence, dislocation association at high temperatures, that is enabled by the shallow energy landscape may lead to embrittlement in SrTiO\textsubscript{3} when type $\mathcal{C}$ dislocations are the most relevant type.

        As discussed above, climb dissociation further immobilizes a dislocation.
        Because climb dissociation usually starts from a full dislocation, we next turn to evaluating the possibility of climb dissociation of type $\mathcal{C}$ dislocations.

        While glide dissociation happens on the \hkl{110} plane along a \hkl<1-10> direction -- as known from type $\mathcal{A}$ dislocation, see Ref. \cite{Hirel2010} -- climb dissociation would imply a stacking fault on a \hkl{11-2} plane.
        This stacking fault, however, is not feasible.
        Imagine the shift in the stacking fault plane for the case of splitting into two equal collinear partials.
        In the glide stacking fault on a \hkl{110} plane the basic sequence of layers is retained and cation-containing \hkl{110} planes are still separated by oxygen ion \hkl{110} planes.
        On the contrary, the stacking fault on a \hkl{11-2} plane brings cations much closer together.
        Neither in any experiment known to the authors nor in our simulations such a stacking fault could be found or artificially created.
        This is fundamentally different from type $\mathcal{A}$ dislocations where the stacking fault plane for glide and climb are crystallographically equivalent.

        \paragraph{\textbf{Loading the dislocation}}
        Based on the experience from type $\mathcal{A}$ dislocations, we first focus on the charge balanced and glide dissociated type $\mathcal{C}$ dislocation in the loading simulation.
        Because this dislocation type is of mixed character, the applied load to drive dislocation motion should also have more than one component.
        We distinguish two scenarios.
        In the first scenario we apply $\tau_{xy}$ to drive dislocation motion via the edge part, in the second scenario the screw part of the mixed dislocation reacts to $\tau_{xz}$.
        The obtained Peierls stresses for low temperature and short dislocation lines are in the ranges of \qty{5.00}{\giga\pascal} -- \qty{5.25}{\giga\pascal} and \qty{2.75}{\giga\pascal} --\qty{3.00}{\giga\pascal}, respectively.

        Taking a more detailed look at the microscopic picture we recognize that dislocations emit an array of defects at the original location of the dislocation prior to moving.
        Therefore, there seems to be a force pinning the dislocation.
        Additionally, dislocations move in very curved fashion and emit many vacancies and vacancy arrays during their motion.
        Analog to the above observations where we observed that dislocations are only mobile in their charged state we suspect that this array of defects acts to restore the original charged dislocation configuration.
        Therefore, the values measured for $\tau^m_{xy}$ and $\tau^m_{xz}$ are probably too high.

        Consequently, we perform the same loading tests with the as-created charged dislocations, like displayed in the left part of \Cref{fig:type_C_charge_imbalance_variants}.
        These types of dislocations can be easily moved using stresses of $\tau^m_{xy} \approx \qty{750}{\mega\pascal}$ and $\tau^m_{xz} \approx \qty{400}{\mega\pascal} - \qty{450}{\mega\pascal}$ at \qty{10}{\kelvin} in the smaller simulation cells.
        If kink formation is enabled by studying sufficient dislocations line length (60 unit cells) these values even reduce to \qty{600}{\mega\pascal} -- \qty{650}{\mega\pascal} and \qty{350}{\mega\pascal} -- \qty{400}{\mega\pascal}, respectively.
        Additionally, the glide of this dislocation type does not produce any point defects in the wake of the dislocation line and could, thus, be sustained for longer glide distances.
        
        These low values for the Peierls barrier are quite surprising.
        Compared to type $\mathcal{A}$ dislocations, a type $\mathcal{C}$ dislocations can be expected to enable much larger plastic deformation per dislocation.
        At the moment it is not clear why the Peierls barrier is reduced so strongly or why the charged dislocations seems stabilized during glide.

        \paragraph{\textbf{Outlook}}
        Concluding, a loss of ductility by climb dissociation of type $\mathcal{C}$, as is expected for type $\mathcal{A}$ dislocations, is not probable.
        While type $\mathcal{A}$ dislocations can be rendered sessile by their climb dissociation, this process is unfeasible for type $\mathcal{B}$ and $\mathcal{C}$.
        From a standpoint of plasticity, a $\mathcal{C}$-type dislocation is, therefore, considered more potent than a type $\mathcal{A}$ dislocation.

        We confirmed the large contribution of type $\mathcal{C}$ dislocations to plasticity in loading simulations.
        Here we could show that the glide dissociated configuration is more stable under load than the pure edge dislocation type while also the Peierls barrier is reduced.
        Only when given enough time, a system containing this dislocation type will equilibrate and dislocations can passivate by collecting or emitting oxygen vacancies.
        This results in more or less sessile dislocations and prevents plastic deformation.
        
        \begin{table}[ht] 
            \caption{Qualitative overview of the mobility of dislocation types $\mathcal{A}$, $\mathcal{B}$, and $\mathcal{C}$.}
            \label{tab:mobility_overview}
            \begin{center}
                \begin{tabular}{r | c c}
                    \toprule
                    type & as-created & charge balanced \\
                    \midrule
                    $\mathcal{A}$ & mobile & sessile\\
                    $\mathcal{B}$ & \textit{is charge balanced} & mobile\\
                    $\mathcal{C}$ & mobile & sessile\\
                    \bottomrule
                \end{tabular}
            \end{center}
        \end{table}

        In \Cref{tab:mobility_overview} we give a short qualitative overview of the results obtained so far.
        Next, we will relate this dislocation type $\mathcal{C}$ to another observed structure in plastically deformed SrTiO\textsubscript{3}.

    \subsection{Type $\mathcal{D}$ dislocation} 
      
        This dislocation type with $\vec{b} = a$\hkl<110> has been observed in TEM studies, and it was proposed that it has a tendency to dissociate by climb \cite{Jin2013, Porz2021}.
        However, in contrast to the edge dislocation $\mathcal{A}$ with line vector \hkl<001> this dislocation $\mathcal{D}$ has a line vector of \hkl<1-11>, see \Cref{fig:types_overview}.
        Its glide plane would be a \hkl{1-1-2} plane and climb dissociation would mean a dissociation splitting along a \hkl<1-1-2> direction in a \hkl{110} plane.
        These glide planes are familiar, see above.
        Consequently, we do expect a possibility for this dislocation to form a stable climb dissociated form but no stable glide dissociation.

        If glide dissociation and, thus, high mobility of this dislocation type is not expected, then how can a type $\mathcal{D}$ dislocation be introduced in an SrTiO\textsubscript{3} crystal?
        Jin \emph{et al.} suggest that the reaction of two type $\mathcal{C}$ dislocations is the source \cite{Jin2013}.
        Such a reaction can indeed happen according to the following reaction between two antiparallel mixed dislocations where the screw part cancels. A first type $\mathcal{C}$ dislocation,
            \begin{alignat}{3}
                &\vec{b}^{(1)} &&= \vec{b}_{\text{edge}}^{(1)} &&+ \vec{b}_{\text{screw}}^{(1)} ,\\
                &a \hkl<110> &&= \frac{a}{3} \hkl<11-2> &&+ \frac{2a}{3} \hkl<111> ,
            \end{alignat}
        and a second type $\mathcal{C}$ dislocation,
            \begin{alignat}{3}
                &\vec{b}^{(2)} &&= \vec{b}_{\text{edge}}^{(2)} &&+ \vec{b}_{\text{screw}}^{(2)} ,\\
                &a \hkl<-10-1> &&= \frac{a}{3} \hkl<-12-1> &&+ \frac{2a}{3} \hkl<-1-1-1> ,
            \end{alignat}
        interact:
            \begin{alignat}{4}
                &\vec{b}_{\text{edge}}^{\text{(new)}} &&= \frac{a}{3} \hkl<11-2> &&+ \frac{a}{3} \hkl<-12-1> &&= a \hkl<01-1> ,\\
                &\vec{b}_{\text{screw}}^{\text{(new)}} &&= \frac{2a}{3} \hkl<111> &&+ \frac{2a}{3} \hkl<-1-1-1> &&= \vec{0} .
            \end{alignat}

        Nevertheless, such a reaction produces a full edge dislocation of type $\mathcal{D}$.
        Climb dissociation is then still a matter of thermal activation or activation energy introduced by electron irradiation in e.g. transmission electron microscopes \cite{Porz2021}.
        In the following we, therefore, test the equilibrium structure of this dislocation type as well as the behavior under long annealing procedures.

        \paragraph{\textbf{Simulation}}
        Just as for the other edge dislocations, dislocation dipoles are created by a cut-and-remove procedure.
        Again, following the above recipes, the charge balancing is checked first.
        As-created dislocations carry an excess charge because the cut planes are non-stoichiometric, see \Cref{fig:type_D_charge_imbalance}.
        The charge imbalance can again be compensated by shifting of half of the oxygen ions from one dislocation to the other.
        Alongside the balancing of charge this also lowers the total potential energy of the simulation cell due to reduction of electrostatic energy contributions.

\begin{figure}[htb]
            \centering
            \includegraphics[]{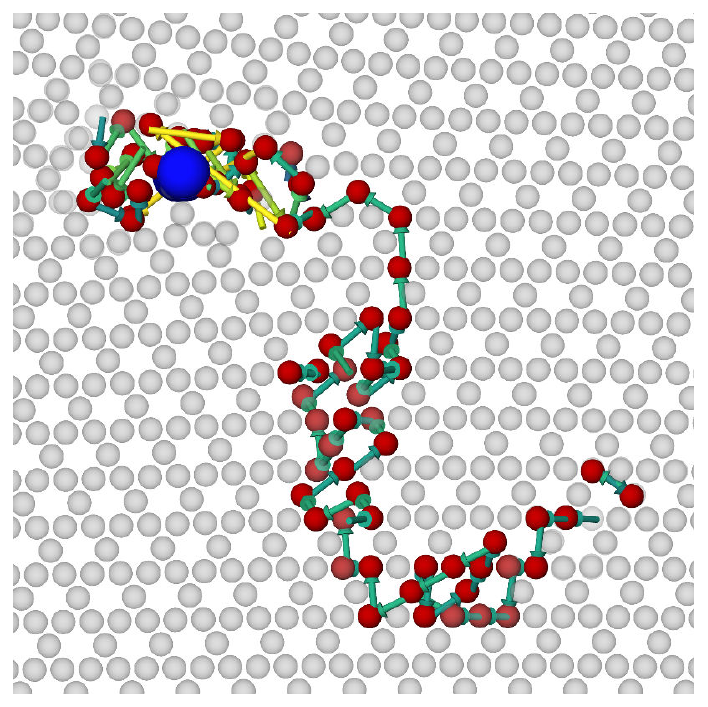}
            \caption{Type $\mathcal{D}$ dislocation after long annealing at high temperature. Only oxygen ions are shown for clarity and color coding in red represents position where the oxygen vacancy has visited. The dislocation Burgers vector is still localized as a full dislocation (represented in blue). One oxygen vacancy has been emitted from the dislocation core and diffused away which reduced the positive dislocation core charge.}
            \label{fig:type_D_annealed_Ox_migration}
        \end{figure}

        First, glide dissociation does not occur for this dislocation spontaneously.
        Instead, the full Burgers vector remains localized in a full dislocation, irrespective of the dislocation core charges.
        Even during simulations where annealing at \qty{3000}{\kelvin} over several hundred picoseconds has been applied, the dislocation does neither dissociate by glide nor climb.
        When the dislocations are not charge balanced, however, we find that the positively charged dislocations emit oxygen vacancies after long annealing periods.
        Thereby the charge at the dislocation core is reduced, see \Cref{fig:type_D_annealed_Ox_migration}.

        For even longer equilibration timespans at elevated temperatures we, thus, expect that the charge at the dislocations equilibrates.
        To continue the investigation, the charge balanced configuration is, consequently, assumed to be the physically stable state.
        As glide dissociation does not occur, climb dissociation is tested analog to \Cref{fig:type_A_energies}.

        \begin{figure*}[htb] 
            \centering
            \includegraphics[]{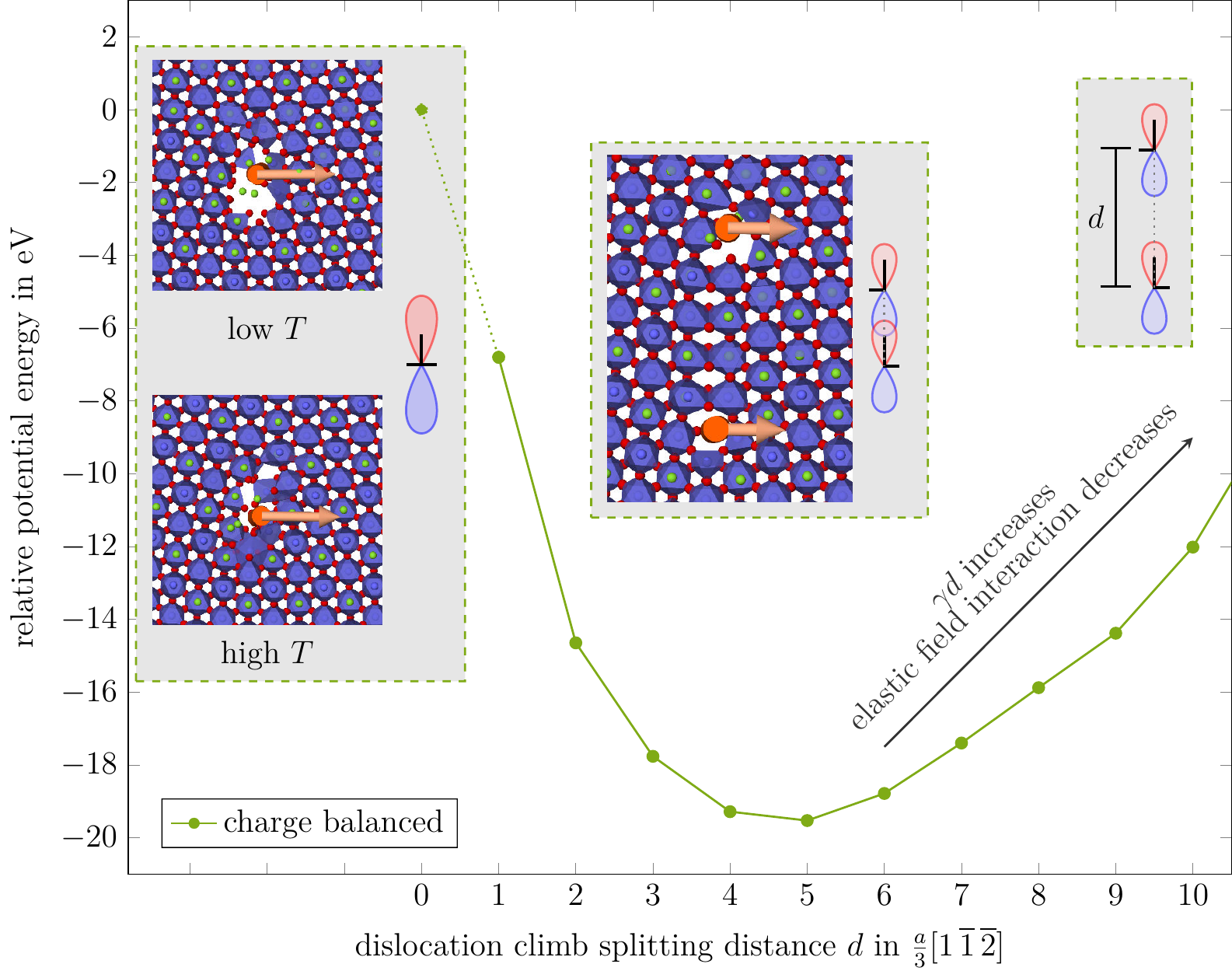}
            \caption{Comparison of energies of type $\mathcal{D}$ dislocations for different climb splitting distances. Each point represents the average of at least three independent samples. Just the charge balanced results are shown which behave identical to the non charge balanced models.}
            \label{fig:type_D_energies}
        \end{figure*}

        Similar to the finding for the other edge dislocation type a climb dissociation lowers overall potential energy, see \Cref{fig:type_D_energies}.
        From an energetic point of view the climb dissociated form is, thus, the stable configuration.
        Significant activation energy, however, prevents this dissociation from happening even at very high temperatures.

        \paragraph{\textbf{Outlook}}
        To sum up, the creation of a type $\mathcal{D}$ dislocation as the product of a dislocation reaction seems to be the only feasible way.
        Compared with the previously discussed dislocation types it is assumed to be rather immobile due to the large and undissociated Burgers vector.
        Additionally, climb dissociation further reduces its mobility once sufficient diffusion is possible.
        In the light of macroscopic plasticity of SrTiO\textsubscript{3} type $\mathcal{D}$ dislocations probably play a minor role.
        For this reasons moving the dislocation under mechanical load has not been attempted.

    \subsection{Type $\mathcal{E}$ dislocation}

        To complicate matters, there is evidence that the slip system with Burgers vector \hkl<110> and slip plane \hkl{1-10} is not the only one that can be activated at room temperature.
        Using nanoindentation experiments on polycrystalline SrTiO\textsubscript{3} Mao and Knowles showed that also a type $\mathcal{E}$ dislocation, being of mixed type with $\vec{b} = \hkl<110>$ but a line vector \hkl<100> and a \hkl{001} slip plane, can be active \cite{Mao1996}.
        Note that they have a slip plane different from the previously discussed dislocation types, see \Cref{fig:types_overview}.
        This work was supplemented with theoretical explanations by Yang \emph{et al.} \cite{Yang2009}.
        Apparently, these dislocations have been found to appear mostly in mixed (line orientation \hkl<100>), but also as pure edge and pure screw dislocations.
        While the earlier of the two references claims that the mixed dislocations appear in glide as well as climb dissociated form after sintering, the latter reference finds dislocation dipoles as a prominent feature.
        It is suggested that these edge dipoles appear when two antiparallel edge dislocations pass each other and get trapped by their elastic interactions while the observed screw dipoles are an artifact of sessile edge dislocation jogs that are bowed out under stress.

        Additionally, Yang \emph{et al.} point out that with five available slip systems (2 x \hkl<110>\hkl{1-10} and 3 x \hkl<110>\hkl{001}) SrTiO\textsubscript{3} in principle satisfies the Taylor criterion and \textquote{significant plasticity in polycrystalline SrTiO\textsubscript{3} ceramics can be expected} \cite{Taylor1938}.
        Note however, that the complex stress states with very high applied stresses in indentation can produce all kinds of dislocations since fracture on the nanoscale is inhibited and Peierls barriers can be easily overcome.
        This interpretation is supported by the very high Peierls barrier that is expected for this dislocation type \cite{Ferre2008}.
        Nevertheless, the prospect of a possible further slip system motivates us to also perform a (rudimentary) investigation of this dislocation type.

        \paragraph{\textbf{Glide and climb}}
        In simulation, this dislocation can be introduced into SrTiO\textsubscript{3} crystal slabs similar to type $\mathcal{C}$ dislocations.
        Some further complications arise for this dislocation type compared to the other ones discussed before.
        
\begin{figure*}[htb] 
            \centering
            \includegraphics[]{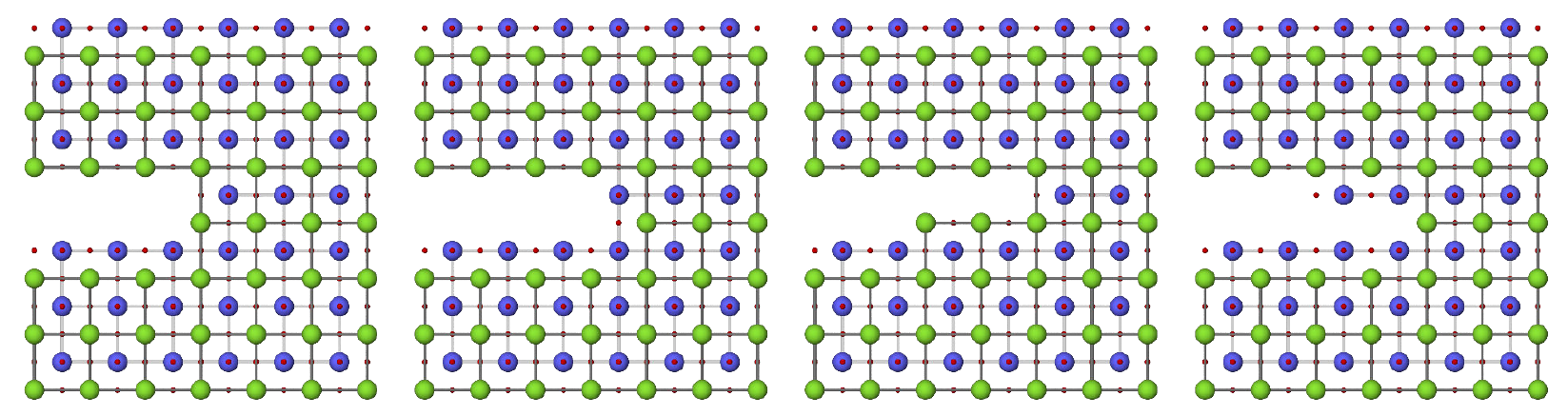}
            \caption{
                Different configurations of type $\mathcal{E}$ dislocations.
                Two pictures on the left: Dislocation can be either terminated by an SrO (green) or by a TiO\textsubscript{2} layer (blue).
                Two pictures on the right: A climb stacking fault is either constituted by two adjacent SrO layers of by two adjacent TiO\textsubscript{2} layers.
                }
            \label{fig:type_E_variants}
        \end{figure*}

        First, the cut for the edge contribution of this dislocation can be made in two different ways.
        The removed plane of atoms can either be terminated by an SrO or by a TiO\textsubscript{2} layer, see \Cref{fig:type_E_variants}.
        Although both layers are nominally charge balanced, it can be expected that they behave differently.
        This problem has also been encountered when studying dislocations with Burgers vector \hkl<100>, line vector \hkl<001> and slip plane \hkl{010} \cite{Metlenko2014}.

        Second, the reports of the splitting of such type $\mathcal{E}$ dislocations in Ref. \cite{Mao1996} suggest that the stacking fault energy hypersurface or $\gamma$-surface is not trivial and possesses at least a local minimum in between full Burgers vector shifts.
        We have already calculated and discussed the $\gamma$-surface in \Cref{fig:sf_surface}.
        The local minimum apparent in the $\gamma$-surface of the \hkl{1-10} is responsible for the stable splitting of dislocations with \hkl{1-10} glide plane, such as dislocations of type $\mathcal{A}$.
        In contrast, the \hkl{100} $\gamma$-surface is rather simple and no intermediate local minima appear.
        A splitting of type $\mathcal{E}$ dislocations into glide dissociated partials can, thus, be ruled out.
        
        Third, climb dissociation has also been reported \cite{Mao1996}.
        For climb dissociation of either dislocation termination there are again two possibilities, see \Cref{fig:type_E_variants}.
        In one case the atomic planes in the climb stacking fault are arranged in a way that a layer of SrO is removed, i.e. the stacking fault consists of two directly neighboring TiO\textsubscript{2} layers.
        In the other case a TiO\textsubscript{2} is removed and now two SrO layers are placed directly adjacent to each other.
        Again, the energetic situation for both climb dissociations can be expected to differ.
        For the additional screw component of this dislocation type the behavior during splitting is not immediately obvious.

        However, the possibility to study this dislocation type using the given interatomic potential is limited.
        The interatomic potential by Thomas \emph{et al.}, was created using the ionic charge as a fit parameter only requiring the full SrTiO\textsubscript{3} formula unit to be charge neutral \cite{Thomas2005}.
        Therefore, the strontium and titanium oxide layers are actually layers of SrO\textsuperscript{0.44} and TiO\textsubscript{2}\textsuperscript{-0.44}.
        Only if nominal charges were assumed, i.e., Sr\textsuperscript{2+}, Ti\textsuperscript{4+}, and O\textsuperscript{2-}, the terminating layers and stacking fault surfaces in \Cref{fig:type_E_variants} would all be charge neutral.
        This is not of concern for the above dislocation types and generally gives good results for the dislocation behavior.
        Yet, in the case of pure SrO and TiO\textsubscript{2} surface terminations of type $\mathcal{E}$ dislocations the charge imbalance will lead to unrealistically high electrostatic repulsion.
        Consequently, the climb dissociated configurations of this dislocation type cannot be realistically studied with this method.
        Also, the cores of the SrO and TiO\textsubscript{2} terminated dislocations are slightly charged (\Cref{fig:type_E_variants} left).

        \begin{figure}[htb]
            \centering
            \includegraphics[]{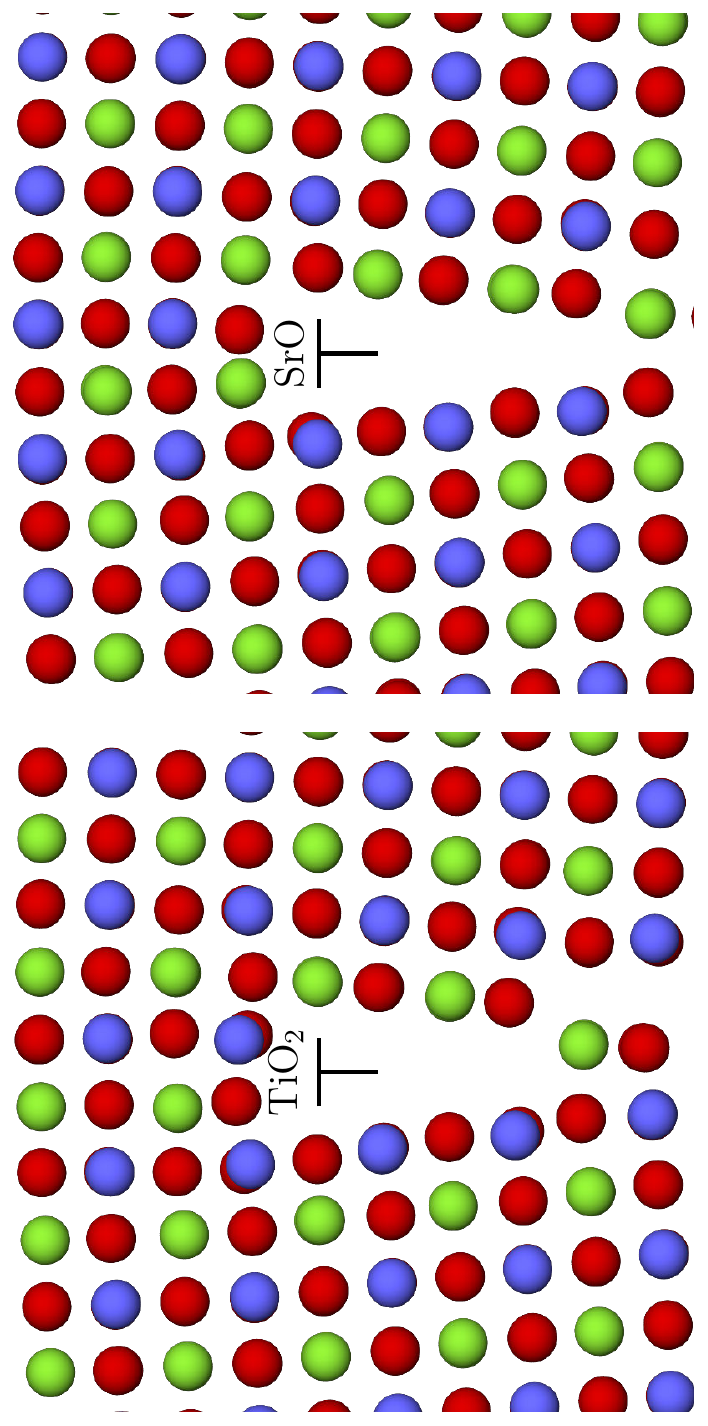}
            \caption{Type $\mathcal{E}$ dislocation with SrO terminal layer (top) and TiO\textsubscript{2} terminal layer (bottom). Neither of the two dislocation types show significant rearrangement.}
            \label{fig:type_E_relaxed}
        \end{figure}

        \paragraph{\textbf{Simulation}}
        Keeping in mind these fundamental restriction we, nevertheless, find it worthwhile to at least model the full dislocations with different dislocation cores (not ensuring exact stoichiometry in this particular case).
        In simple relaxations at very low temperatures the dislocation structure of either SrO or TiO\textsubscript{2} terminated dislocation cores shows no rearrangement, see \Cref{fig:type_E_relaxed}.
        Spontaneous dislocation splitting is, thus, not occurring.
        In order to test for thermally activated glide dissociation of this dislocation type again temperatures of \qty{2500}{\kelvin} are applied to the system.
        Even after long annealing periods the dislocation structure remains essentially unchanged except for some slight disordering around the dislocations core.
        Neither glide nor climb dissociation can be observed.
        Thus, it is expected that this type $\mathcal{E}$ dislocation cannot lower its Peierls barrier by dislocation splitting.
        This finding is consistent with the calculated $\gamma$-surfaces.
        Concluding, type $\mathcal{E}$ dislocation are expected to play a minor role in macroscopic plasticity of SrTiO\textsubscript{3}.

\section{Summary}

    After carefully reviewing existing literature on mechanically introduced dislocation in strontium titanate (SrTiO\textsubscript{3}) with Burgers vector $a$\hkl<110>, it has become clear that the dislocation structure is still a matter of intensive debate.
    To clarify the equilibrium configuration of the many dislocation types proposed and search for hints of their influence on plasticity, all the prominent dislocation types have been modeled atomistically.
    By addressing the dislocation equilibrium structure we reveal important hints regarding mobility and multiplication, which are crucial ingredients for ceramic plasticity.

    It was found that only dislocations with a \hkl{1-10} glide plane (types $\mathcal{A}$, $\mathcal{B}$ and \& $\mathcal{C}$) are easy to glide by mechanical stress due to the possible glide dissociated configuration.
    Other glide planes such as the \hkl{001} glide plane (type $\mathcal{E}$) prohibit this mechanism of Peierls barrier reduction or prevent glide altogether, as is the case for the \hkl{1-1-2} glide plane (type $\mathcal{D}$).
    As suspected by Mitchell \emph{et al.}, the ionic nature of SrTiO\textsubscript{3} forces the stacking faults to retain some stacking sequence that does not bring ions of equal sign charge too close together.
    Therefore, it determines which dislocation may split into partial dislocations. 

    Especially for the studied dislocations with (partial) edge character (type $\mathcal{A}$ \& $\mathcal{C}$) we discovered that the ionic nature of SrTiO\textsubscript{3} needs to be accounted for when studying dislocation structures.
    These dislocation cores are inherently charged but can be charge balanced by changing the oxygen occupancy at the dislocation core \cite{Whitworth1975,Marrocchelli2015}.
    The tendency of dislocations to dissociate into partials was shown to be very sensitive to this dislocation core charge state, and thus possibly very sensitive to the relevant point defect equilibria and point defect mobility.
    Since, in turn, dislocation mobility is strongly related to the dissociation of dislocations, the macroscopic ductility is directly related to the charge at the dislocation core.
    Yet the effect is qualitatively different from the well known pinning by charged clouds that accumulate at or around a dislocations \cite{Eshelby1958}.
    
    Additionally, we confirmed that the most intensively studied dislocation type $\mathcal{A}$ is inherently meta-stable with the full Burgers vector allocated in a single dislocation as well as distributed in a glide dissociated configuration.
    By carefully preparing and comparing glide and climb dissociated dislocation arrangements it was shown that the climb dissociated configuration is stable.
    On the one hand, a loss of dislocation mobility is associated with dislocation climb dissociation while, on the other hand, dislocation climb is a known precondition for dislocation multiplication mechanisms such as the multiple cross glide mechanism (cf. Gilman and Johnston \cite{Gilman1962}).
    Whether dislocation immobilization by climb dissociation or enabling multiplication is the dominant effect when it comes to macroscopic plasticity in SrTiO\textsubscript{3} remains to be shown.
    
    Ultimately, we deem dislocations of type $\mathcal{B}$ and $\mathcal{C}$ to be most important for plasticity in SrTiO\textsubscript{3}.
    Their Peierls barrier is low and they do not have a tendency to immobilize by climb dissociation.

    In general, we found very good agreement between the results of our simulations and analytical considerations on the basis of linear elastic theory, the ionic charges, and the crystal structure.
    This knowledge alone allowed making very useful predictions on dislocations in SrTiO\textsubscript{3}.
    An important point that is elusive to these considerations and has been left for further studies is the interaction of the dislocation structure and point defect equilibria.
    Founded on the observed changes to dislocations structure with the change of the oxygen ion concentration at the dislocation core, significant changes may be expected for more complex compositions.

\section*{Acknowledgments}
    Calculations for this research were conducted on the Lichtenberg high performance computer of the TU Darmstadt.
    The authors also acknowledge helpful discussions with Prof. Dr. J\"urgen R\"odel in the course of preparing the publication.
    Financial support to Arne J. Klomp has been granted by the Deutsche Forschungsgemeinschaft via the SPP 1599.
    Lukas Porz is greatly indebted to the Deutsche Forschungsgemeinschaft for funding under grant number 398795637 and 414179371.

\bibliography{bib/complete}

\clearpage
\appendix

\section{Charge density for type $\mathcal{D}$ dislocations}

\begin{figure}[htbp]
    \centering
    \includegraphics[]{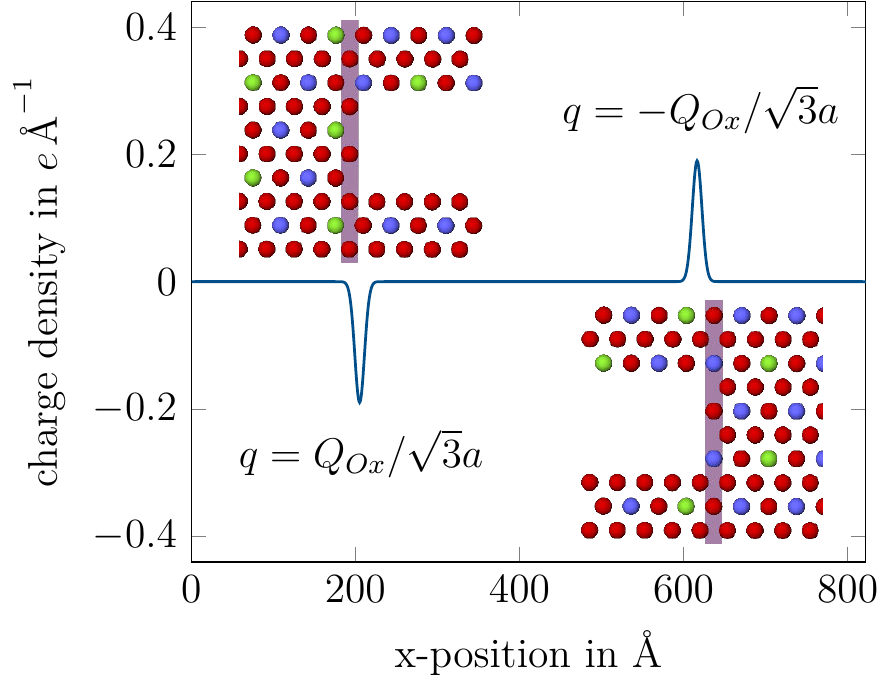}       
    \caption{
        Charge density in a simulation cell containing a type $\mathcal{D}$ dislocation dipole.
        In order to obtain a smooth charge density that can be binned along the x-direction, the ionic point charges have been smeared out be a Gaussian distribution.
        Additionally, the integrated charges $q$ for each dislocation and a representation of the terminal layer of the dislocation cut are given.
        }
    \label{fig:type_D_charge_imbalance}
\end{figure}

\end{document}